\documentclass[aps,prb,amsmath,amssymb,twocolumn]{revtex4}

\usepackage{graphicx}
\usepackage{adjustbox}
\usepackage{bm}
\usepackage{color}
\usepackage{braket}
\usepackage{standalone}
\usepackage{multirow}
\usepackage{tikz}
\usepackage{mathrsfs}
\usepackage{dsfont}
\usepackage[colorlinks,bookmarks=true,citecolor=blue,linkcolor=red,urlcolor=blue]{hyperref}

\newcommand{\bea}{\begin{eqnarray}}
\newcommand{\eea}{\end{eqnarray}}
\definecolor{dgreen}{rgb}{0.1,0.5,0.1}
\definecolor{red}{rgb}{1,0,0}

\usepackage{bbm}
\usepackage[caption=false]{subfig}
\usepackage{floatrow}
\usepackage{stackrel}
\usepackage{wrapfig}

\usepackage{tikz-feynman}
\usepackage{tikz}
\usetikzlibrary{positioning}

\newcommand{\tikzmark}[1]{\tikz[overlay,remember picture] \node (#1) {};}
\newcommand{\DrawBox}[4][]{%
    \tikz[overlay,remember picture]{%
        \coordinate (TopLeft)     at ($(#2)+(-0.2em,0.9em)$);
        \coordinate (BottomRight) at ($(#3)+(0.2em,-0.3em)$);
        \path (TopLeft); \pgfgetlastxy{\XCoord}{\IgnoreCoord};
        \path (BottomRight); \pgfgetlastxy{\IgnoreCoord}{\YCoord};
        \coordinate (LabelPoint) at ($(\XCoord,\YCoord)!0.5!(BottomRight)$);
        \draw [red,#1] (TopLeft) rectangle (BottomRight);
        \node [below, #1, fill=none, fill opacity=1] at (LabelPoint) {#4};
    }
}

\begin{document}

\title{Exceptional Non-Hermitian Phases in Disordered Quantum Wires}
\author{Benjamin Michen}
\author{Tommaso Micallo}
\author{Jan Carl Budich}
\email{jan.budich@tu-dresden.de}
\affiliation{Institute of Theoretical Physics${\rm ,}$ Technische Universit\"{a}t Dresden and W\"{u}rzburg-Dresden Cluster of Excellence ct.qmat${\rm ,}$ 01062 Dresden${\rm ,}$ Germany}
\date{\today}

\begin{abstract}
We demonstrate the occurrence of nodal non-Hermitian (NH) phases featuring exceptional degeneracies in chiral-symmetric disordered quantum wires, where NH physics naturally arises from the self-energy in a  disorder-averaged Green's function description. Notably, we find that at least two nodal points in the clean Hermitian system are required for exceptional points to be effectively stabilized upon adding disorder. We identify and study experimental signatures of our theoretical findings both in the spectral functions and in mesoscopic quantum transport properties of the considered systems. Our results are quantitatively corroborated and exemplified by numerically exact simulations on a microscopic lattice model. The proposed setting provides a conceptually minimal framework for the realization and study of topological NH phases in quantum many-body systems.
\end{abstract}

	\maketitle
\section{Introduction}
Novel phases of matter combining the paradigm of topological insulators and semimetals \cite{HasanKane2010,Qi2011,Wen2017,Armitage2018} with the notion of effective non-Hermitian (NH) Hamiltonians describing dissipative systems \cite{BreuerPetruccione,Ashida2020} have become a broad frontier of current research \cite{Rudner2009, Zeuner2015, lee,Xu2017, Zhou2018, Lieu2018, Longhi2018, BBC, yaowang, Imhof2018, gong, Kawabata2019, Budich2019, Yoshida2019, EPringExp, KawabataEP2019, KunstDwivedi2019, Song2019, QdynBBC, SzameitScience2020, Budich2020, review, Feinberg1997_Ref_B, Mondragon2013_Ref_B, Ostahie2021_Ref_B}.
 With applications ranging from the field of classical robotic meta-materials \cite{Brandenbourger2019,Ghatak2020} to the realm of quantum many-body systems \cite{Kozii2017,Bergholtz2019,EPManyBody,McClarty2019,Rausch2021}, various microscopic mechanisms of dissipation have been found capable of inducing such NH topological phases (see Ref.~\cite{review} for an overview). 

A prominent example in the context of quantum materials is provided by impurity scattering in disordered systems. Beyond simply giving a finite life-time to the Bloch states of a clean solid, disorder potentials exhibiting a non-trivial structure in internal degrees of freedom such as spin may induce a NH self-energy $\Sigma$ with a complex matrix structure. In particular, if $\Sigma$ does not commute with the free Hamiltonian $H_0$, topologically stable NH semimetal phases may be stabilized, as has been predicted in both two-dimensional and three-dimensional systems by means of perturbative calculations 
\cite{EPDisorder1, EPDisorder2, EPDisorder3}. Generally speaking, in NH systems the notion of band-touching points such as Dirac- and Weyl-nodes is generically replaced by exceptional points (EPs) \cite{BerryDeg, Heiss, EPreview}, i.e. degeneracies at which the NH Hamiltonian becomes non-diagonalizable. 

 \floatsetup[figure]{style=plain,subcapbesideposition=top} 
\begin{figure}[htp!]	 
  \centering 
  \sidesubfloat[]{\includegraphics[trim={-1.7cm 7.7cm 1.7cm 0cm}, width=0.95\linewidth]{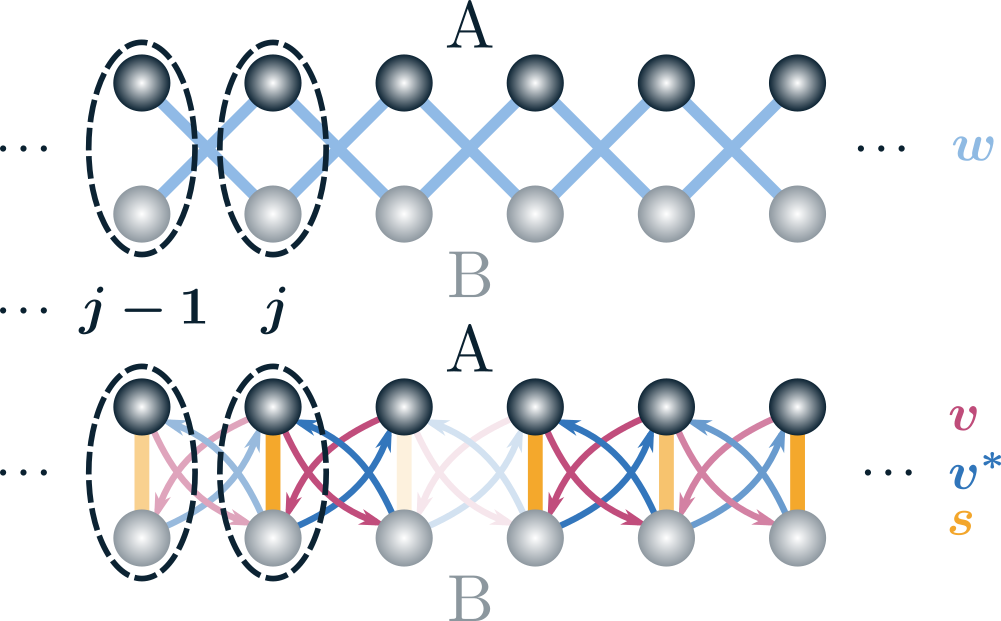} \label{fig1a}} \\
  \sidesubfloat[]{\includegraphics[trim={-1.7cm 0cm 1.7cm 0cm}, clip, width=0.95\linewidth]{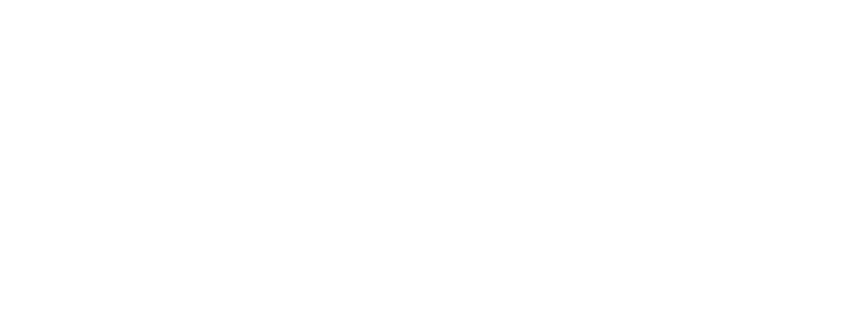} \label{fig1b} }\\ 
  \sidesubfloat[]{\includegraphics[trim={0cm 0cm 0cm 0cm}, clip, width=0.95\linewidth]{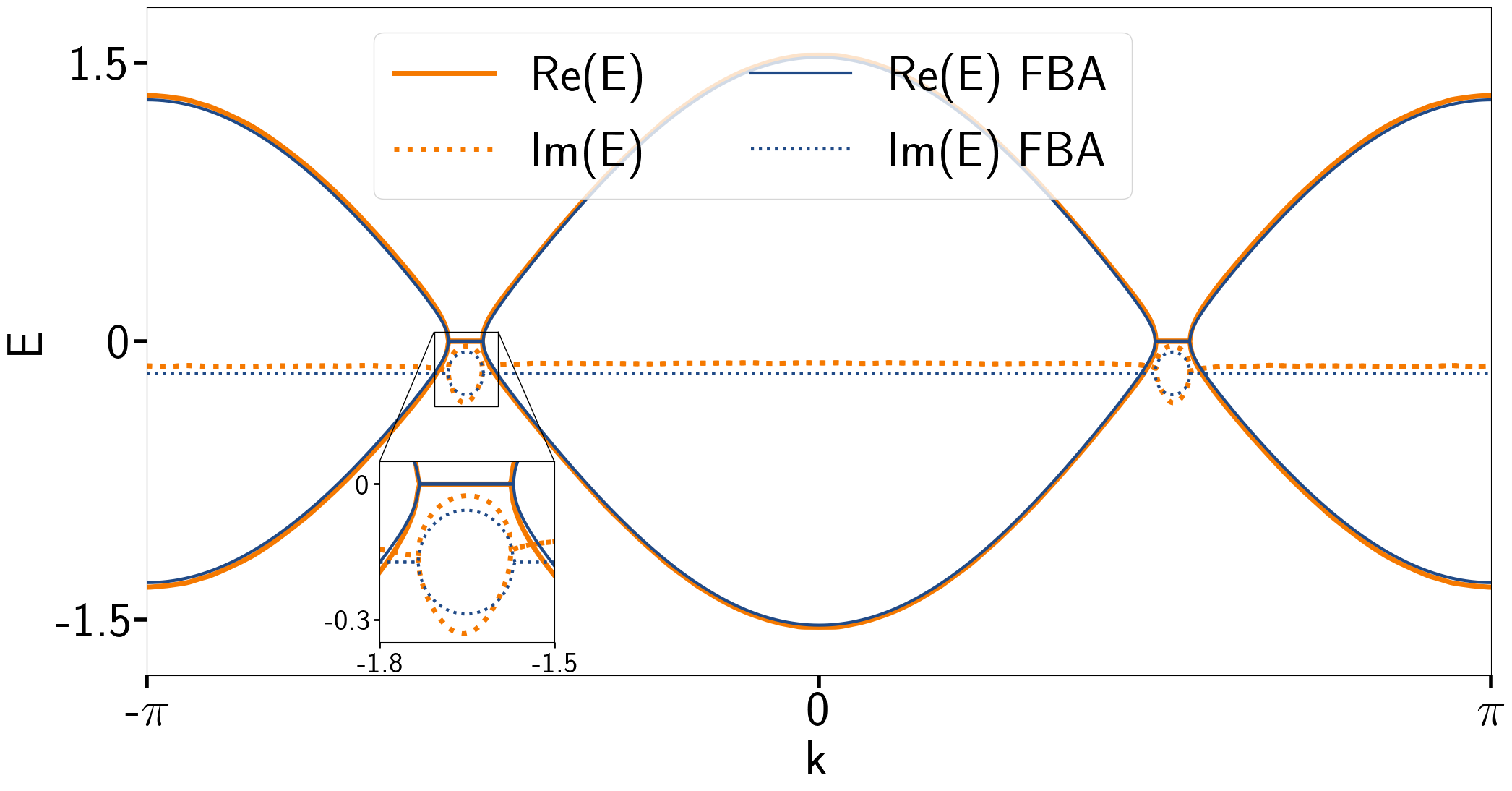} \label{fig1c} }
\caption{(a) Illustration of the translation-invariant model  Hamiltonian $H_0$ (see Eq. (\ref{H_0})).
(b) Illustration of the disorder term $V$ (see Eq.~(\ref{disorder})).  $V$ consists of random amplitude {\color{black} intracell} and nearest-neighbour terms. 
(c) Complex spectrum of the effective NH Hamiltonian $H_e$ (see Eq.~(\ref{eqn:heff})) of a disordered system microscopically described by the total Hamiltonian $H = H_0 + V$ (see Eqs.~(\ref{H_0}-\ref{disorder})). Exact numerical results are shown in orange, perturbative results within first born approximation in blue. Two pairs of exceptional points connected by Fermi arcs are visible (see also inset). Parameters are  { \color{black}  $w = -1 / \sqrt{2}$, $\alpha =0.7$, $s = (1 + i) / \sqrt{2}$, $v=-0.5 (1 + i) / \sqrt{2}$. }
 }
\end{figure}

Here, we report on the discovery of topologically stable exceptional NH phases in disordered {\emph{one-dimensional}} quantum systems with chiral symmetry. To this end, we study the effective NH Hamiltonian
\begin{align}
H_{e}(k) = H_0(k) + \Sigma(k,\omega=0),
\label{eqn:heff}
\end{align}
associated with the disorder averaged Green's function \cite{MBQT} at the Fermi Energy $(\omega = 0)$ that gives rise to the NH self-energy $\Sigma$. While we find that a single Hermitian nodal point in $H_0$ cannot split into a pair of stable EPs to leading order, impurity scattering between two band-touching points is shown to yield a NH band structure with four EPs (see Fig.~\ref{fig1c} for an illustrating example). We study the physical properties of this intriguing NH nodal phase. Specifically, basic observables such as spectral functions and the linear response conductance in a simple two-terminal mesoscopic quantum transport setting \cite{Datta} are compared to a conventional disordered phase without EPs, thus pinpointing experimentally accessible differences between the inequivalent NH phases. Our results are qualitatively derived within a perturbative approach and quantitatively corroborated by means of numerically exact simulations.\\

\section{Microscopic model}
We propose and study a microscopic two-band model in one spatial dimension (1D) that is described by the Hamiltonian $H = H_0 + V$, consisting of a translation-invariant part $H_0$ (see Fig.~\ref{fig1a} for an illustration) that is perturbed by a random disorder term $V$ (see Fig.~\ref{fig1b}). Specifically,
\begin{align}
H_0 =& \sum_{j = 1} ^ {N-1} w \left(\psi_{j+1,B}^\dagger\psi_{j,A} + \psi_{j,B}^\dag\psi_{j+1,A}\right)  + \mathrm{h.c.},  \label{H_0}
\end{align}
where length is measured in units of the lattice constant and energy in terms of the nearest neighbor hopping chosen as { \color{black} $w = -1 / \sqrt{2}$}, the field-operators $\psi_{j,A (B)}$ annihilate a fermion in unit cell $j$ on sublattice  A (B) (cf.~Fig.~\ref{fig1a}). Assuming periodic boundary conditions, the corresponding Bloch Hamiltonian in reciprocal space takes the form $H_0(k) = {\bm d_R(k) \cdot  \bm \sigma}$ with { \color{black}  $ {\bm d}_R(k) =  - \sqrt{2} (\mathrm{cos}(k), 0, 0)$}, and
{ \color{black} obeys the symmetry  $[H_0,\sigma_x] = 0$, i.e. the A-B sublattice pseudo-spin on which the standard Pauli-Matrices $\bm \sigma$ act has a conserved quantization axis along the $x$-direction}. The spectrum of $H_0(k)$ exhibits 
two { \color{black} nodal points at $k=\pm \frac{\pi}{2}$, which are crucial for the emergence of EPs, and do not depend on our specific choice of $w$.} The random disorder term $V$ is similarly 
structured (cf.~Fig.~\ref{fig1b}) and has the real space representation 
\begin{align}
V =&  \sum_{j = 1} ^ {N-1} a_j \bigg [s\,  \left(\psi_{j,A}^\dagger \psi_{j,B} + \mathrm{h.c.}\right) \nonumber \\
&+ v  \left(\psi_{j+1,B}^\dagger\psi_{j,A} + \psi_{j,B}^\dag\psi_{j+1,A}\right) + \mathrm{h.c.}  \bigg], \label{disorder}
\end{align}
where $s , v \in \mathbb C$, and the amplitudes $\{ a_j \}$ are uncorrelated and drawn from the uniform distribution on the real interval $[- \alpha, \alpha]$, i.e. the overall disorder strength is parameterized by $\alpha$. It is worth noting that the correlation in amplitude between the {\color{black} intracell}  and nearest-neighbor terms that share the same random amplitude $a_j$ is important for the formation of EPs. {\color{black} Thus, as we demonstrate below, a nodal NH phase with EPs or a gapped phase arises depending on the parameters $s$ and $v$ in (\ref{disorder}). We emphasize that the disorder-term $V$ only preserves a residual chiral symmetry $\sigma_z V \sigma_z = - V$ but breaks the aforementioned higher symmetry of $H_0$. That a more fragile Hermitian nodal phase is promoted to an exceptional NH phase only requiring a lower symmetry \cite{Bergholtz2019} reflects and exemplifies the lower co-dimension (or higher stability) of EPs as compared to diagonalizable gap-closing points in Hermitian systems \cite{review}.}

To obtain the effective Hamiltonian $H_e(k) = H_0(k) + \Sigma(k,\omega=0)$ (cf.~Eq.~(\ref{eqn:heff})), we calculate  the retarded Green's function in frequency space, i.e. the Fourier transform  of  the propagator $G^\mathrm{R}_{k,k'}(t-t') =-i\Theta(t-t') \langle 0|\{ c_k(t), c^\dagger_{k'}(t')\}|0 \rangle$ with respect to $(t-t')$, where  
$c_k^\dag(t) = (c_{k,A}^\dag(t), c_{k,B}^\dag(t))$ denotes the Heisenberg picture A-B sublattice spinor of creation operators in reciprocal space. Upon averaging over the impurity amplitudes, translational invariance is restored, thus rendering the resulting disorder averaged Green's function $G^{\mathrm{R,av}}_{k}$ block diagonal in momentum space \cite{MBQT}. From
$G^{\mathrm{R,av}}_{k}(\omega) = [\mathbbm{1}(\omega + i \eta) - H_0(k) - \Sigma(k,\omega)]^{-1}$, where $\eta >0$ is an infinitesimal regularization, we then infer the self-energy correction $\Sigma(k,\omega)$ and the effective Hamiltonian $H_e(k)$, respectively.\\

\begin{figure}[htp!]
  \centering
  \sidesubfloat[]{\includegraphics[trim={0cm 0cm 0cm 0cm},clip, width=0.95\linewidth]{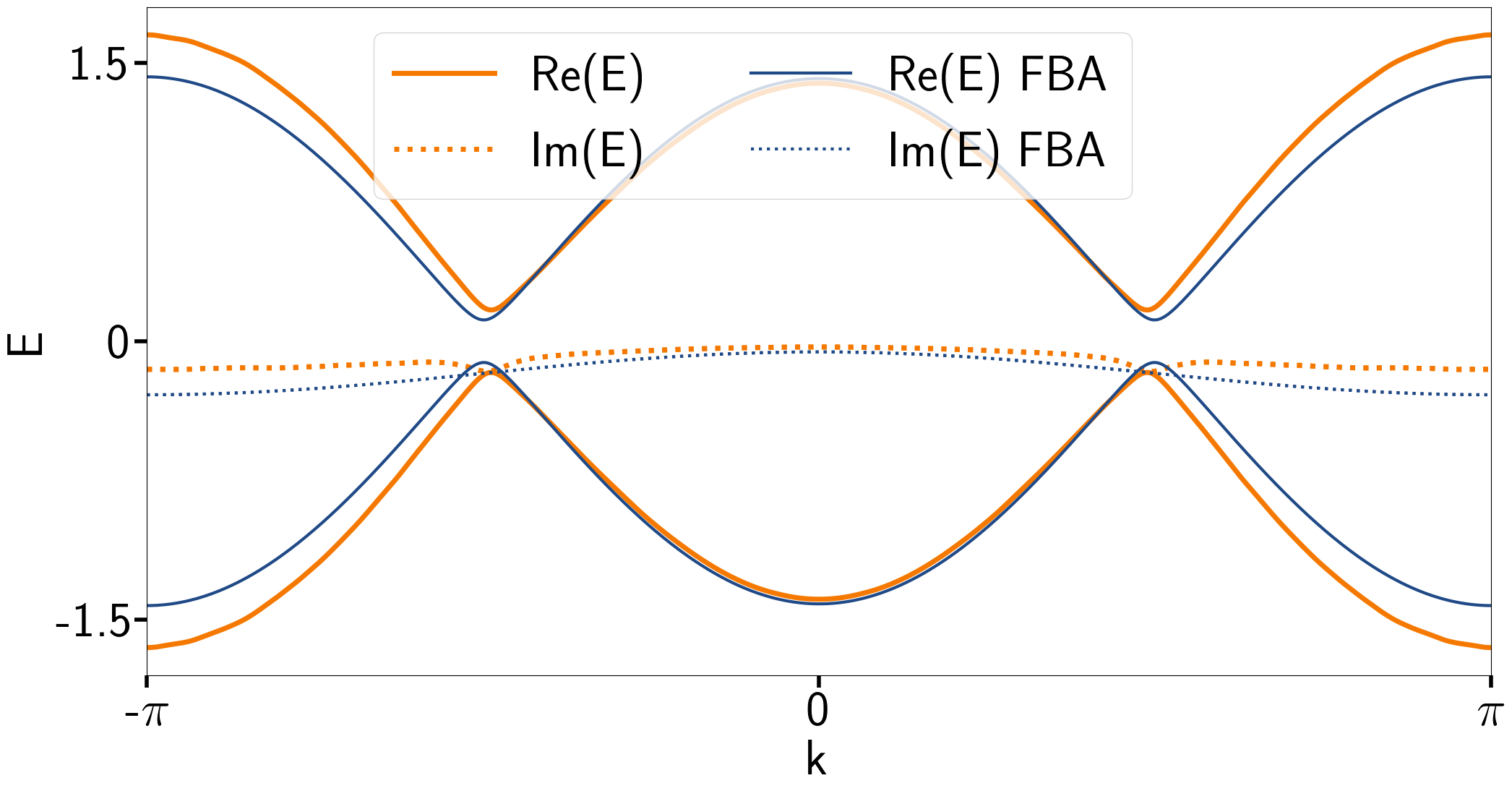} \label{Spec_No_EP}}\\
  \sidesubfloat[]{\includegraphics[trim={1.5cm 1cm 6cm 2.5cm}, clip,
width=0.95\linewidth]{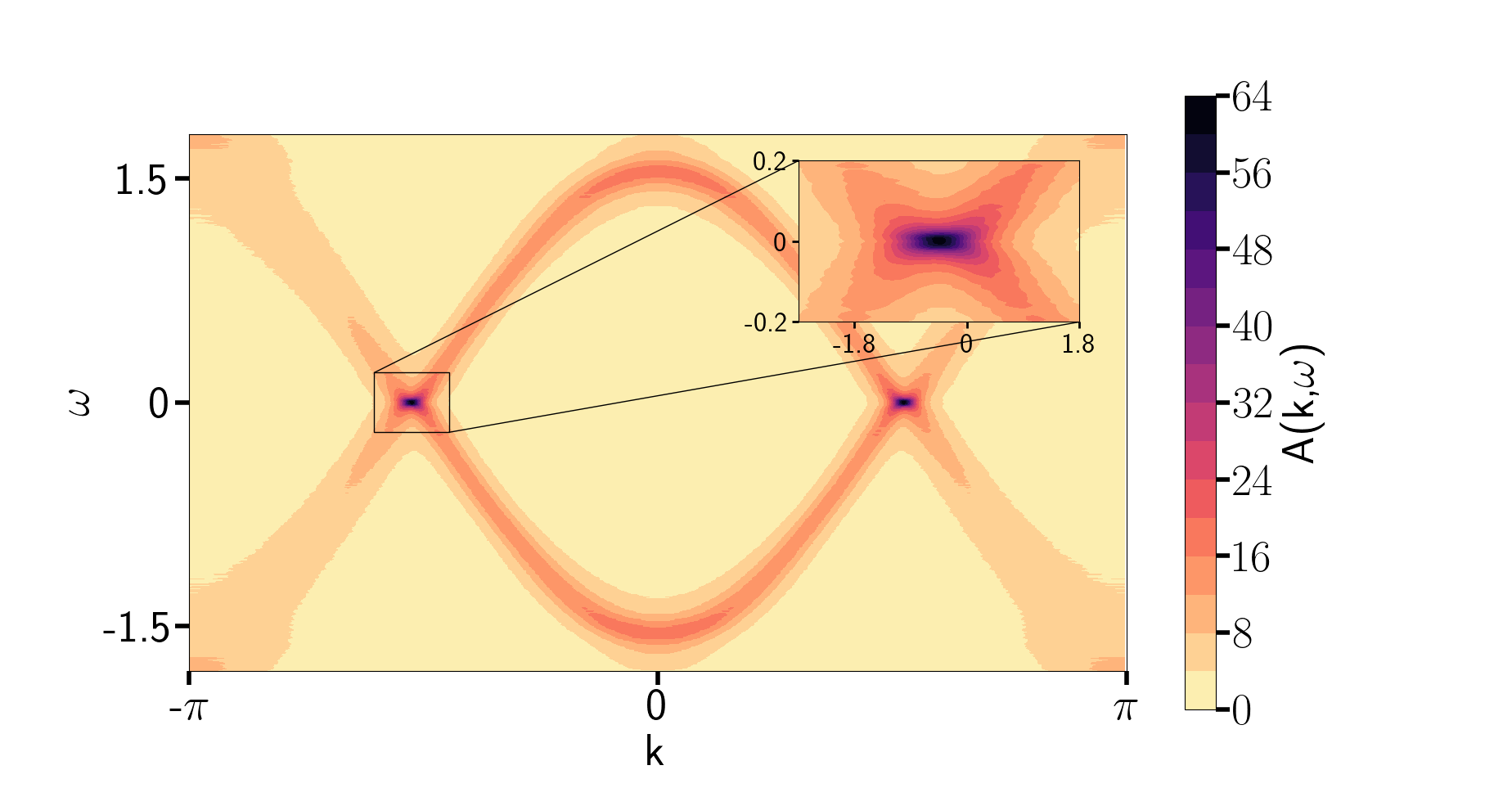} \label{Spec_Func_EP} }\\
\sidesubfloat[ ]{\includegraphics[trim={1.5cm 1cm 6cm 2.5cm}, clip,  width=0.95\linewidth]{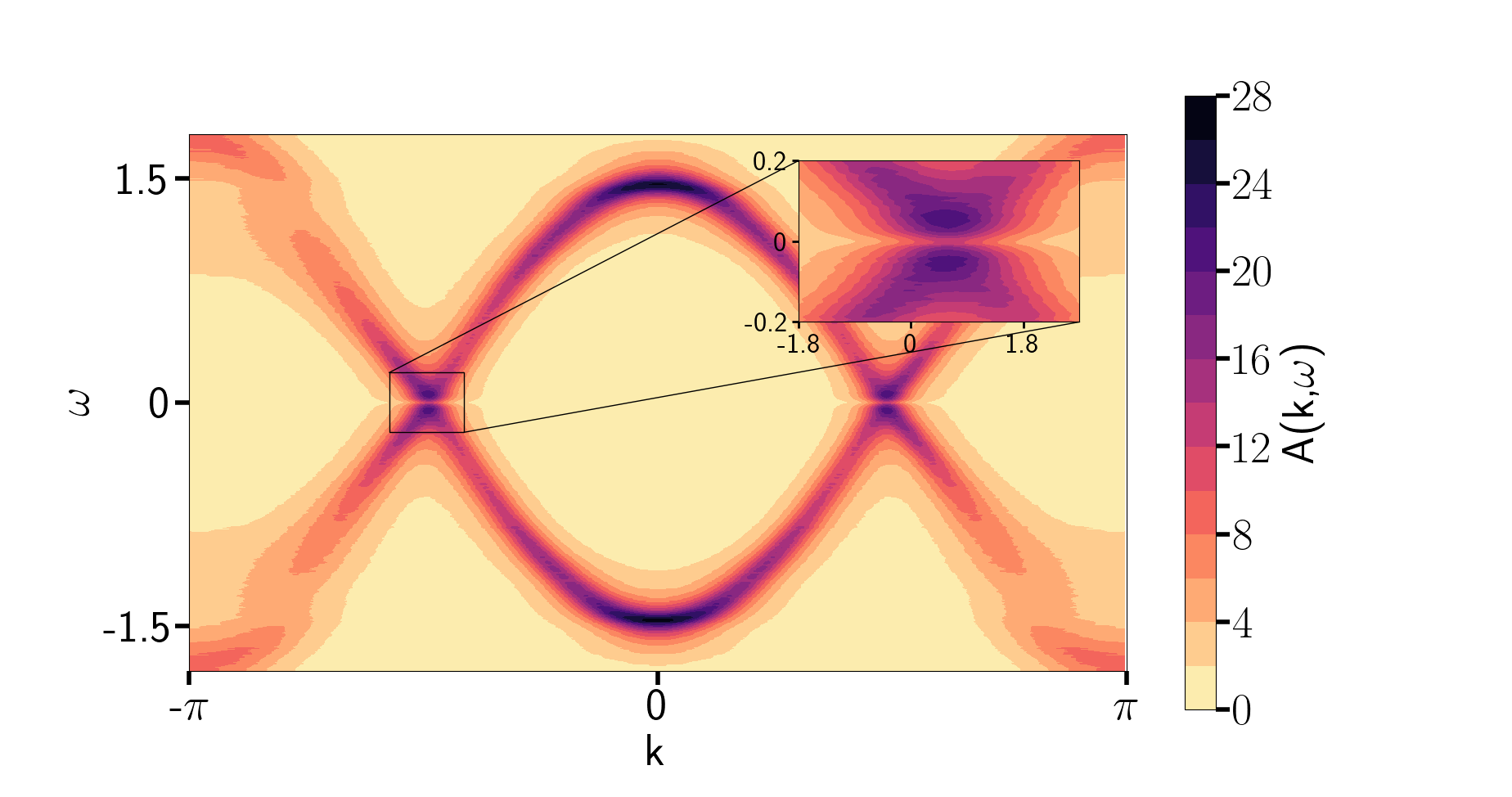} 
\label{Spec_Func_No_EP}}
\caption{(a) Spectrum of $H_e(k)$ in the conventional phase. The exact numerical result is plotted in orange, the FBA result is shown in blue for comparison.
Parameters are  { \color{black}  $\alpha =0.7$, $s = (1 + i) / \sqrt{2}$, $v=-0.5 (1 -i) / \sqrt{2}$.} (b) Exact numerical result for the spectral function  $A(k,\omega)$ of the exceptional phase. Parameters are  { \color{black}  $\alpha =0.7$, $s = (1 + i) / \sqrt{2}$, $v=-0.5 (1 + i) / \sqrt{2}$}. (c) Exact numerical result for the spectral 
function  $A(k,\omega)$ of the conventional phase. Parameters are  { \color{black} $\alpha =0.7$, $s = (1 + i) / \sqrt{2}$, $v=-0.5 (1 - i) / \sqrt{2}$.}}
\end{figure} 
\section{Disorder-induced exceptional degeneracies}

To investigate the occurrence of disorder-induced EPs in our model system (\ref{H_0}-\ref{disorder}), it is convenient to represent the effective NH Hamiltonian (\ref{eqn:heff}) in the form
\begin{align}
H_e(k) = d_0(k) \sigma_0 + \mathbf{d}(k)\cdot \boldsymbol{\sigma},
\end{align}
where $\mathbf{d} = \mathbf{d}_R+ i \mathbf{d}_I$ with $\mathbf{d}_R, \mathbf{d}_I \in \mathbb R^3$ is the complex generalization of a Bloch vector, and $d_0 \in \mathbb C$ is a complex energy shift. The degeneracy points of the complex spectrum $E_\pm = d_0 \pm \sqrt{\mathbf{d}_R^2 - \mathbf{d}_I^2 + 2 i \mathbf{d}_R \cdot \mathbf{d}_I}$ of $H_e$ generically represent EPs and rely on two real conditions that amount to a vanishing real and imaginary part under the latter square-root function. Due to chiral symmetry, $d_0$ must be purely imaginary and the imaginary part of the aforementioned EP conditions, namely $\mathbf{d}_R \cdot \mathbf{d}_I =0$, is always satisfied. The only remaining condition for an EP is then given by
\begin{align}
\mathbf{d}_R^2 = \mathbf{d}_I^2
\label{eqn:epcond}
\end{align}
In our model system, Eq.~(\ref{eqn:epcond}) is typically satisfied at pairs of momenta where the Hermitian part of $H_e$ is close to an ordinary diagonalizable nodal point (i.e. close to $\lvert \mathbf d_R\rvert =0$), and an anti-Hermitian term $\sim i \sigma_z$ associated with the symmetry allowed $z$-component of $\mathbf d_I$ occurs in the self-energy (cf.~Eq.~(\ref{eqn:heff})). 
In this sense, the self-energy $\Sigma$ induced by the disorder term (\ref{disorder}) may split Hermitian gap-closing points into pairs of EPs. However, as we analytically derive further below, for this mechanism to be effective at least a pair of Hermitian degeneracy points must be present in the unperturbed spectrum, thus leading to a minimum of four EPs in the complex energy band-structure of $H_e$.

Our numerically exact results on the complex spectrum of the disorder-averaged effective Hamiltonian $H_e$ (see Eq.~(\ref{eqn:heff})) are compared to first Born approximation (FBA) calculations in Fig.~\ref{fig1c} and Fig.~\ref{Spec_No_EP}, respectively.
 For the parameter choice  { \color{black} $s = (1 + i) / \sqrt{2}$, $v=-0.5 (1 + i) / \sqrt{2}$} in Eq.~(\ref{disorder}), a nodal NH phase featuring two pairs of EPs, at which both the real and the imaginary part of the effective energy become degenerate, is stabilized (see  Fig.~\ref{fig1c}). The underlying disorder-free Hamiltonian $H_0$ (see Eq.~(\ref{H_0})) has ordinary band crossings at $k=\pm \pi / 2$.
 We note that the distance between the EPs occurring around each of those crossings at weak disorder continuously increases with the overall disorder strength $\alpha$. Furthermore, a so called NH Fermi-arc, i.e. a continued two-fold degeneracy in the real part of the spectrum of $H_e$, connects the EPs. The splitting of this degeneracy from the EPs exhibits a characteristic square-root dispersion, in contrast to the at least linear splitting of degeneracies in the Hermitian realm.

By contrast, for the parameter choice  { \color{black} $s = (1 + i) / \sqrt{2}$, $v=-0.5 (1 -i) / \sqrt{2}$} a gapped phase without EPs is observed (see Fig.~\ref{Spec_No_EP}). There, anti-crossings open around the nodal points of the unperturbed Hamiltonian $H_0$, and the corresponding energy gap increases with the overall disorder strength $\alpha$.
 
 \section{Experimental signatures}
From the analysis of the effective Hamiltonian $H_e$ (see Eq.~(\ref{eqn:heff})) that provides phenomenological insights about the quasi-particle excitations close to the Fermi energy, we now turn to identifying more immediately observable characteristics of the disorder-induced exceptional NH phase. Specifically, we start by comparing the spectral function $A(k,\omega) = -2 \mathrm{Im}(\mathrm{Tr}[G^{\mathrm{R,av}}_{k,k}(\omega)])$ for the exceptional phase
(see Fig.~\ref{Spec_Func_EP}) to the conventional phase in (see Fig.~\ref{Spec_Func_No_EP}). Most notably, the spectral function of the exceptional phase peaks at zero energy along the Fermi arcs (cf. Fig.~\ref{fig1c}), whereas the spectral function of the conventional phase almost vanishes at zero energy and exhibits two ridges to the right and left, which is indicative of the spectral gap in Fig.~\ref{Spec_No_EP}. We note that the exceptional phase may also readily be distinguished from the disorder-free system which is hallmarked by band-crossings at isolated points rather than extended Fermi-arcs. 

\begin{figure}[htp!]
  \centering
  \includegraphics[trim={0cm 0cm 0cm 0cm},clip, width=\linewidth]{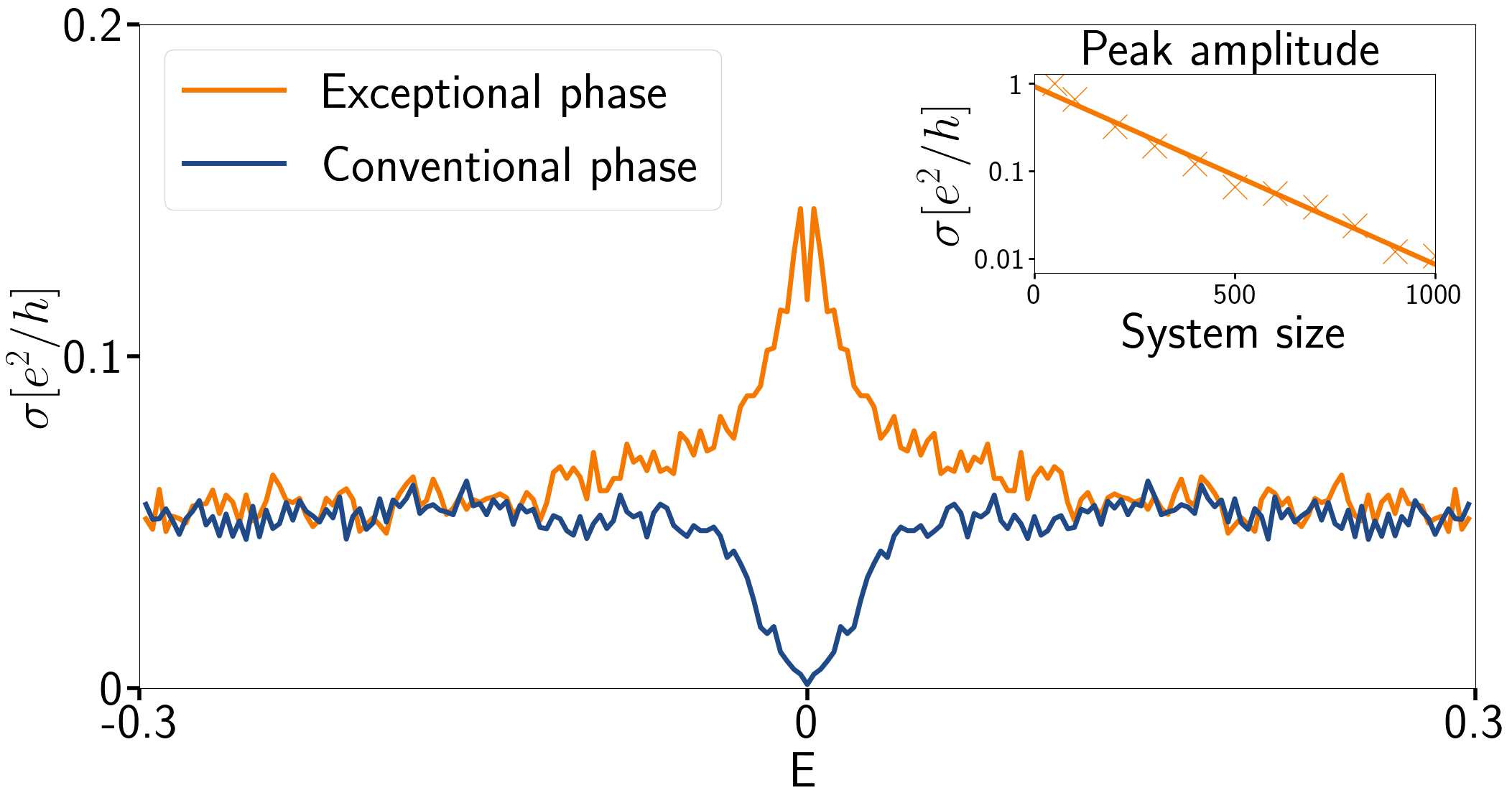}
\caption{Energy-dependent two-terminal conductance of a disordered system (cf. Eqs.~(\ref{H_0}-\ref{disorder})) with $N= 400$ sites in the exceptional and conventional phase (see plot legend). The result is averaged over 1000 disorder realizations. The parameters in the exceptional phase are  { \color{black}  $\alpha =0.3$, $s = (1 + i) / \sqrt{2}$, $v=-0.5 (1 + i) / \sqrt{2}$ } and in the ordinary phase   { \color{black} $\alpha =0.3$, $s = (1 + i) / \sqrt{2}$, $v=-0.5 (1 -i) / \sqrt{2}$.} Inset: Amplitude of the zero-energy conductance peak (averaged over the energy interval [-0.015, 0.015]) in the exceptional phase as a function of system size in a semi-log plot along with an exponential 
fit $\sigma(l) = \sigma_0 \mathrm{e}^{-\mathrm{ln}( 2) l / l_0 }$ with $\sigma_0 = 0.93 e^2/h$, $l_0 = 147.9$.}
\label{Transmission_peak_inset}
\end{figure}

As a second experimental signature, we study the linear-response conductance in a simple two-terminal mesoscopic quantum transport setting 
{\color{black} by  calculating the transmission amplitude in the standard $S$-matrix formalism} \cite{Datta}. The thermal reservoirs (leads) attached to both ends of the system are modeled with the same dispersion as the unperturbed Hamiltonian $H_0$. In Fig.~\ref{Transmission_peak_inset}, we compare the transmission as a function of energy for the exceptional and the gapped phase through a system with 400 sites at disorder strength $\alpha =0.3$. The exceptional phase shows a pronounced peak around zero energy whereas the gapped phase exhibits a strong suppression of transport in that energy region. This constitutes a qualitative difference between the two inequivalent disorder-induced NH phases occurring in our model system. Yet, the transport properties of the exceptional phase are distinct from the metallic disorder-free phase that would exhibit a conductance of $2 e^2/h$ that does not decay with system size. By contrast, due to the small but finite imaginary part of the zero-energy states constituting the Fermi-arcs (cf.~Fig.~\ref{fig1c}), the zero-energy conductance of the exceptional phase exhibits a slow exponential decay with system size (see inset of Fig.~\ref{Transmission_peak_inset}). The peak amplitude for each system length is taken as the average of the transmission over the energy interval [-0.015, 0.015]. An exponential fit reveals a decay length of about 150 sites. Our transport simulations are performed using the Kwant library \cite{Kwant}.

\section{General analysis of disorder-induced EPs}
Our unperturbed model Hamiltonian $H_0$ (see Eq.~(\ref{H_0})) features two band-crossing points at the momenta $k = \pm \pi/2$. It is natural to ask whether this represents a minimal setting or whether it is also possible to split a single nodal point into a pair of EPs. We now derive on general grounds that indeed at least two nodal points are required to obtain EPs from leading order disorder scattering in one-dimensional systems. This result is {\emph{independent}} of the microscopic details of $H_0$, especially its number of bands. Our analysis is based on perturbation theory in FBA, and qualitatively agrees with all our data from exact numerical simulations of two banded model systems.The main steps of our argumentation are as follows.

We first derive an analytical expression for the anti-Hermitian (AH) part of the self-energy in FBA. In the case of disorder with random amplitudes, the self-energy correction in FBA generally takes the form

\begin{align}
\Sigma(k, \omega) = \langle a \rangle_f V_{k,k} + \frac{1}{N} \langle a ^2 \rangle_f \sum_{p} V_{k,p}G^{R,0}_{p} ( \omega) V_{p,k}, \label{SE_FBA}
\end{align}
where $\langle ... \rangle_f $ denotes the expectation value with respect to the probability distribution $f$ of the amplitudes $\{ a_j \}$,  $V_{k,p}$ the matrix-valued disorder scattering vertex in reciprocal space, and $G^{R,0}_{p} ( \omega)$ the free retarded Green's function. As any matrix, the self-energy can be uniquely decomposed into a Hermitian and an AH part as $\Sigma(k, \omega)  = \Sigma^{\mathrm{H}}(k, \omega)  + \Sigma^{\mathrm{AH}}(k, \omega)$. In the continuum limit of Eq. (\ref{SE_FBA}), the AH part can be written as 

\begin{align}
&\Sigma^{\mathrm{AH}}(k, \omega)  =  -\frac{i}{2} \langle a ^2 \rangle_f  \sum_{m=1}^{n}\sum_{k_m \in K_m}  \prod_{\substack{j =1\\ j \neq m}}^n  
\frac{1}{\omega - E_j(k_m(\omega))} \nonumber \\
 &\times \frac{1}{\lvert E_m'(k_m(\omega)) \rvert }   V_{k,k_m(\omega)}\mathrm{adj}[\mathbbm{1}\omega - H_0(k_m(\omega))]V_{k_m(\omega),k}, \label{se_ah}
\end{align}
 where $K_m$ is the set of momenta at which the band $E_m(k)$ of the the free Hamiltonian $H_0(k)$ intersects the energy $\omega$, $n$ the number of bands, and adj[...] denotes the adjugate matrix. We derive this result in Appendix \ref{Analytical_AH_SE}. Roughly speaking, this equation tells us that there is a term contributing to $\Sigma^{\mathrm{AH}}(k, \omega)$ for each momentum at which a band crosses the energy $\omega$.

Using this intuition, we now investigate a system with a single Dirac point at zero energy. We assume that two bands $E_l$ and $E_{l+1}$ cross at some momentum $k_\mathrm{t}$ with a 
slope $v = \lvert E'_l(k_\mathrm{t})\rvert = \lvert E'_{l+1}(k_\mathrm{t})\rvert$, where the latter equality sign follows from chiral symmetry. We then take the limit $\omega \to 0$ of Eq. (\ref{se_ah}) by linearizing
$E_l(k_\mathrm{t} + q)  = -v q$, $E_{l +1}(k_\mathrm{t} + q)  = v q$ and calculate
$\Sigma^{\mathrm{AH}}(k, \omega =  0 ) = \lim_{q \to 0} \Sigma^{\mathrm{AH}}(k, \omega =  vq )$. In the vicinity of the nodal point, we can effectively describe the system as a two-band model by projecting it onto the subspace spanned by the eigenstates of the two crossing bands $\lvert E_l (k)\rangle$ and $\lvert E_{l+1} (k)\rangle$. Marking the projected operators by a tilde, the linearized effective model reads as  $\widetilde{H}_0(k_\mathrm{t} - q) = vq \sigma_z$. The projection of the self-energy directly at the nodal point $k_\mathrm{t}$ is then found to take the form

\begin{align}
\widetilde{\Sigma}^{\mathrm{AH}}(k_\mathrm{t}, \omega = 0) =-\frac{i}{2v} \langle a ^2 \rangle_f (\widetilde{V}_{k_\mathrm{t}, k_\mathrm{t}})^2. \label{Sigma_AH_Projection}
\end{align}
The matrix $\Gamma$ that induces the chiral symmetry by $\Gamma H_0(k)\Gamma^{-1} = - H_0(k)$ can be projected onto the subspace as well. We find 
that $\widetilde{\Gamma}(k_\mathrm{t}) \widetilde{V}_{k_\mathrm{t}, k_\mathrm{t}} (\widetilde{\Gamma}(k_\mathrm{t}))^{-1} = - \widetilde{V}_{k_\mathrm{t}, k_\mathrm{t}}$, which implies that 
$\widetilde{V}_{k_\mathrm{t}, k_\mathrm{t}}$ cannot contain $\sigma_0$ if we represent it as $\widetilde{V}_{k_\mathrm{t}, k_\mathrm{t}} = d_0 \sigma_0 + \mathbf{d}\cdot \boldsymbol{\sigma}$.
In conclusion, $(\widetilde{V}_{k_\mathrm{t}, k_\mathrm{t}})^2$ only contains $ \sigma_0$ and so does $\widetilde{\Sigma}^{\mathrm{AH}}(k_\mathrm{t}, \omega = 0)$
according to Eq. (\ref{Sigma_AH_Projection}).

From this projected two-banded form, we can easily see that the occurrence of EPs is impossible in this setting. If the nodal point of $H_0$ was to be split into a pair of EPs, they would have to be connected by a Fermi-arc with a purely imaginary energy gap. As discussed above, the spectrum of a Matrix $d_0 \sigma_0 + (\bm d_R +  i \bm d_I) \cdot \bm \sigma$ is 
given by $E_\pm= d_0 \pm \sqrt{\bm d_R^2 - \bm d_I^2 + 2 i\bm d_R \cdot \bm d_I}$, and hence $\bm d_R \cdot \bm d_I = 0$ as well as $\bm d_R^2 - \bm d_I^2 <0$ are required along the Fermi-arc. Since $\widetilde{H}_0$ contains $\sigma_z$, $\widetilde{\Sigma}^{\mathrm{AH}}(k_\mathrm{t}, \omega = 0) $ would then have to contain $i \sigma_x$ or $i \sigma_y$ and no $i \sigma_z$ to create a Fermi-arc around momentum $k_\mathrm{t}$ in the spectrum of $\widetilde{H}_{e}(k) = \widetilde{H}_0(k) + \widetilde{\Sigma}(k,\omega=0)$. As we just saw, 
$\widetilde{\Sigma}^{\mathrm{AH}}(k_\mathrm{t}, \omega = 0) $  has a trivial matrix structure, containing only $i \sigma_0$ and thus rendering a Fermi arc impossible.

To sum up, the AH self-energy contribution from disorder cannot lead to any finite splitting of a single nodal point into a pair of EPs within FBA. However, we stress that if a second  nodal point is present, a scattering process between the two points can split them into four EPs (see Appendix \ref{nodal_point_spliting} for a more detailed discussion), as is the case in our model (\ref{H_0}-\ref{disorder}). There, for weak to moderate disorder strengths, the resulting exceptional NH phase is correctly captured within FBA up to small quantitative deviations (see Fig.~\ref{fig1c}).

\section{Concluding discussion}
{\it Concluding discussion.---}
In this work, we have explored the possibility of EPs in disordered one-dimensional systems. To this end, we have presented a two-banded model with two nodal points that can enter an exceptional NH phase through scattering between the nodal points, given a suitable chosen random disorder with up to nearest neighbor terms. A main finding that is valid beyond the studied model system is that a single nodal point cannot split into EPs by potential scattering to leading order.
Furthermore, spectral properties and transport capabilities of the exceptional NH phase are studied and compared to a conventional disordered phase, where the most striking feature of the exceptional phase is an enhanced conductance for energies close to the EPs, which we attribute to zero-energy states with a long lifetime around the Fermi arcs connecting the EPs (cf.~Fig.~\ref{fig1c}). This serves as an experimental distinction between an exceptional and a conventional phase in a chiral-symmetric scenario, where the spectral symmetry $\{E_k\} \to \{-E_k^*\}$ of $H_e$ implies that there is either a gap or a coalescence of eigenvalues (and thus an exceptional degeneracy) at zero energy. The spectral gap of the conventional phase should always result in a transmission gap at zero energy, since there are no states available for transport in the quasiparticle picture. By contrast, the gap closing in the exceptional phase enhances mesoscopic transport at zero energy. Furthermore, even when compared to a conventional disordered system in the absence of chiral symmetry, the anomalously slow decay of the zero-energy conductance with system size is a distinguishing feature of the exceptional NH phases discussed in our present work.

To put our findings into a broader context, we would like to stress a key difference to interaction-induced exceptional NH phases, where EPs result from inter-particle scattering rather than potential scattering \cite{Rausch2021}. There, a finite life-time of quasi-particles that is closely connected to an anti-Hermitian part of the self-energy at the Fermi-surface in thermal equilibrium typically requires finite Temperature. By contrast, all calculations discussed in this work were performed at zero temperature, thus highlighting that disorder-induced NH physics does not rely on the phase-space for scattering processes provided by thermal excitations.

\acknowledgments
{\it Acknowledgments.---}
We would like to thank Emil Bergholtz for discussions. We acknowledge financial support from the German Research Foundation (DFG) through the Collaborative Research Centre SFB 1143, the Cluster of Excellence ct.qmat, and the DFG Project 419241108. Our numerical calculations were performed on resources at the TU Dresden Center for Information Services and High Performance Computing (ZIH).

\onecolumngrid

\appendix

\section{Perturbation Theory}
In this section we will use imaginary timeor  rather imaginary frequency $i \omega$ to shorten the notation. For the final result we will replace $i \omega \to \omega + i \eta$. 
Consider a system consisting of a simple free Hamiltonian $H_0$ plus some perturbation $V$, i.e. $H = H_0 + V$. If $H_0$ has been solved and the free retarded Green's function (GF) $G^0(i \omega) = [\mathbbm{1}i \omega - H_0]^{-1}$ in frequency space is known, a peturbation series for the full retarded GF $G(i \omega) = [\mathbbm{1}i \omega - H]^{-1} = [\mathbbm{1}i \omega - H_0 - V]^{-1}$ arises by self-inserting the Dyson equation as 

\begin{align}
G(i \omega) &= G^0(i \omega)+ G^0(i \omega)V G^0(i \omega)
+  G^0(i \omega)V  G^0(i \omega)V  G^0(i \omega) + ... \label{PTB_series}
\end{align}
If $H_0$ describes a translationally invariant tight-binding model with $n$ internal degrees of freedom on each site,
 the free GF  $G^0_{k,k'}(i \omega) =  \delta_{k,k'} G^0(k, i \omega)$ becomes block-diagonal in the Bloch basis of $H_0$. The blocks of the free GF read $G^0(k, i \omega)=  [\mathbbm{1}i \omega- H_0(k)]^{-1}$, where $H_0(k)$ is the $n$x$n$-Bloch-Hamiltonian.

In our case, the perturbation consists of the same type of impurity at each site but with a random amplitude $a_j$ drawn from a normalized probability distribution $f(a)$, so the impurity part of the Hamiltonian in second quantization reads
\begin{align}
V =& \sum_{j = 1} ^ {N_\mathrm{site}} a_j \left [\frac{1}{2} \Psi_{j}^\dagger V_\mathrm{OS} \Psi_{j} 
+ \Psi_{j+1}^\dagger V_\mathrm{NN} \Psi_{j}+ \Psi_{j+2}^\dagger V_\mathrm{NNN} \Psi_{j}  + ... + \mathrm{h.c.} \right] \nonumber \\
=& \sum_{j= 1} ^ {N_\mathrm{site}} \sum_{k,k'} a_j\frac{e^{-i j (k- k')}}{N_\mathrm{sites}} c_k^\dagger
\underbrace{\left[V_\mathrm{OS} + (V_\mathrm{NN}e^{-ik} + V^\dagger_\mathrm{NN}e^{ik'}) + (V_\mathrm{NNN}e^{-i2k} + V^\dagger_\mathrm{NNN}e^{i2k' }) 
+ ... \right ]}_{V_{k,k'}} c_{k'}. \label{impurity}
\end{align}
The $\Psi_{j}^\dagger$ are $n$-spinors of creators for on-site states and
the $c_k^\dagger = \frac{1}{\sqrt{N_\mathrm{site}}}\sum_j e^{i j k} \Psi_{j}^\dagger$ are $n$-spinors of creators for the Bloch-states. The series from 
Eq. (\ref{PTB_series}) leads to an expression for the blocks of the full GF

\begin{align}
G_{k,k'}(i \omega) =  & \delta_{k,k'} G^0(k, i \omega) + \sum_{j_1=1}^{N_\mathrm{site}}  G^0(k, i \omega)
a_{j_1}\frac{e^{-i j_1 (k - k')}}{N_\mathrm{sites}} V_{k,k'} G^0(k', i \omega) \nonumber \\
& + \sum_{q}\sum_{j_1,j_2 =1}^{N_\mathrm{site}} G^0(k, i \omega)
a_{j_1}\frac{e^{-i j_1 (k - q)}}{N_\mathrm{sites}} V_{k,q} G^0(q, i \omega)
a_{j_2}\frac{e^{-i j_2 (q - k')}}{N_\mathrm{sites}} V_{q,k'} G^0(k', i \omega) +... \nonumber
\end{align}
To obtain the effective Hamiltonian, this expression is averaged over the impurity amplitudes $\{a_j\}$ by calculating
\begin{align}
G_{k,k'}^{av}(i \omega) = \langle G_{k,k'}(i \omega) \rangle_f = \int \mathrm{d}a_1 f(a_1) 
\int \mathrm{d}a_2 f(a_2) ... \int \mathrm{d}a_{N_\mathrm{site}} f(a_{N_\mathrm{site}})
G_{k,k'}(i \omega). \label{G_av} 
\end{align} 

In Eq. (\ref{G_av}) we encounter terms of the form 
$\langle \sum^{N_\mathrm{site}}_{j_1, ...j_m = 1} a_{j_1}a_{j_2}...a_{j_{m}}e^{\sum^m_{l=1} q_l j_l}\rangle_f$. We can group the sums into those where all scattering vectors $q \in \mathbf{Q} = \{q_1, q_2, ..., q_m\}$ are connected to one, two, three and so on impurities. The notation $|\mathbf{Q}_r|$ simply indicates the number of elements in the subset $\mathbf{Q}_r \subset \mathbf{Q} $. 
\begin{align}
\langle \sum^{N_\mathrm{site}}_{j_1, ...j_m = 1} a_{j_1}a_{j_2}...a_{j_{m}}e^{\sum^m_{l=1} q_l j_l}\rangle_f = &
\langle \sum^{N_\mathrm{site}}_{h_1=1} (a_{h_1})^m e^{\sum_{q \in \mathbf{Q}} q h_1}\rangle_f \nonumber \\
&+ \langle \sum_{\cup^2_{r = 1} \mathbf{Q}_r = \mathbf{Q}} \sum^{N_\mathrm{site}}_{h_1=1} 
\sum^{N_\mathrm{site}}_{\substack{h_2 = 1\\ h_2 \neq h_1}}
(a_{h_1})^{|\mathbf{Q}_1|}(a_{h_2})^{|\mathbf{Q}_2|} e^{\sum_{q_1 \in \mathbf{Q_1}} q_1 h_1} 
e^{\sum_{q_2 \in \mathbf{Q_2}} q_2 h_2}\rangle_f \nonumber \\
&+ \langle \sum_{\cup^3_{r = 1} \mathbf{Q}_r = \mathbf{Q}} \sum^{N_\mathrm{site}}_{h_1 = 1} 
\sum^{N_\mathrm{site}}_{\substack{h_2 = 1\\ h_2 \neq h_1}} \sum^{N_\mathrm{site}}_{\substack{h_3 = 1\\ h_3 \neq h_1, h_2}} (a_{h_1})^{|\mathbf{Q}_1|}(a_{h_2})^{|\mathbf{Q}_2|}(a_{h_3})^{|\mathbf{Q}_3|}  \nonumber \\
& \times e^{\sum_{q_1 \in \mathbf{Q_1}} q_1 h_1}
e^{\sum_{q_2 \in \mathbf{Q_2}} q_2 h_2} e^{\sum_{q_3 \in \mathbf{Q_3}} q_3 h_3}\rangle_f \nonumber\\
&+ ... \nonumber
\end{align}
Now we need to introduce a small error of the order $\frac{1}{N_\mathrm{site}}$ by letting the $h$ 
sums run unrestricted, e.g. 
$\sum^{N_\mathrm{site}}_{\substack{h_2 = 1\\ h_2 \neq h_1}} \to \sum^{N_\mathrm{site}}_{h_2 = 1}$. The average doesn't act on the exponentials and the distribution of the amplitudes is uncorrelated, so we can pull the average of the amplitudes out of the sums. Performing the sums gives a delta function for all momenta connected to the same impurity and we arrive at

\begin{align}
\langle \sum^{N_\mathrm{site}}_{j_1, ...j_m = 1} a_{j_1}a_{j_2}...a_{j_{m}}e^{\sum^m_{l=1} q_l j_l}\rangle_f = &
N_\mathrm{site} \langle a^m \rangle_f \delta_{0,\sum_{q \in \mathbf{Q}}q} \nonumber \\
&+ (N_\mathrm{site})^2\sum_{\cup^2_{r = 1} \mathbf{Q}_r = \mathbf{Q} }\delta_{0,\sum_{q_1 \in \mathbf{Q_1}}q_1}
\langle a^{|\mathbf{Q}_1|} \rangle_f \langle a^{|\mathbf{Q}_2|} \rangle_f
 \delta_{0,\sum_{q_2 \in \mathbf{Q_2}}q_2} \nonumber \\
&+ (N_\mathrm{site})^3\sum_{\cup^3_{r = 1} \mathbf{Q}_r = \mathbf{Q}} \langle a^{|\mathbf{Q}_1|} \rangle_f
\langle a^{|\mathbf{Q}_2|} \rangle_f \langle a^{|\mathbf{Q}_3|} \rangle_f
\delta_{0,\sum_{q_1 \in \mathbf{Q_1}}q_1}\delta_{0,\sum_{q_2 \in \mathbf{Q_2}}q_2} \delta_{0,\sum_{q_3 \in \mathbf{Q_3}}q_3} \nonumber \\
&+... \nonumber
\end{align}

Bearing this result in mind, we can express Eq. (\ref{G_av}) in terms of Feynman diagrams. It is given by the sum over all topologically different diagrams of the form

\begin{align}
G_{k,k}^{av}(i \omega)=
\vcenter{\hbox{
\begin{tikzpicture}
  \begin{feynman}
     \vertex (e1) at (-1, 0);
     \vertex (e2) at (0, 0);
     \vertex (i1) at (0, 4)[]{};
     \vertex [above=1em of i1, opacity = 0] {\(\langle a \rangle_f\)};
     \vertex [below=0.5em of e2, opacity = 0] {\(V_{k,k}\)};
    \diagram*{
    (e2) -- [fermion, edge label=\(k\)] (e1),
    (i1) -- [charged scalar, opacity = 0] (e2),
    };
  \end{feynman}
\end{tikzpicture}}}
+
\vcenter{\hbox{
\begin{tikzpicture}
  \begin{feynman}
     \vertex (e1) at (-1, 0);
     \vertex (e2) at (1, 0);
     \vertex (a) at (0, 0);
     \vertex (i1) at (0, 4)[crossed dot]{};
     \vertex [above=1em of i1] {\(\langle a \rangle_f\)};
     \vertex [below=0.5em of a] {\(V_{k,k}\)};
    \diagram*{
    (a) -- [charged scalar, edge label'=\(0\)] (i1),
    (e2) -- [fermion, edge label=\(k\)] (a),
    (a) -- [fermion, edge label=\(k\)] (e1),
    };
  \end{feynman}
\end{tikzpicture}}}
+
\vcenter{\hbox{
\begin{tikzpicture}
  \begin{feynman}
     \vertex (e1) at (-1, 0);
     \vertex (e2) at (2, 0);
     \vertex (a) at (0, 0);
     \vertex (i1) at (0.5, 4)[crossed dot]{};
     \vertex (b) at (1, 0);
     \vertex [above=1em of i1] {\(\langle a^2 \rangle_f\)};
     \vertex [below=0.5em of a] {\(V_{k,q}\)};
     \vertex [below=0.5em of b] {\(V_{q,k}\)};
    \diagram*{
      (b) -- [fermion, edge label=\(q\)] (a) ,
      (a) -- [charged scalar, edge label=\(q-k\)] (i1),
      (b) -- [charged scalar, edge label'=\(k-q\)] (i1),
      (e2) -- [fermion, edge label=\(k\)] (b),
      (a) -- [fermion, edge label=\(k\)] (e1),
    };
  \end{feynman}
\end{tikzpicture}}}
+
\vcenter{\hbox{
\begin{tikzpicture}
  \begin{feynman}
     \vertex (e1) at (-1, 0);
     \vertex (e2) at (2, 0);
     \vertex (a) at (0, 0);
     \vertex (b) at (1, 0);
     \vertex (i1) at (0, 4)[crossed dot]{};
     \vertex (i2) at (1, 4)[crossed dot]{};
     \vertex [above=1em of i1] {\(\langle a \rangle_f\)};
     \vertex [below=0.5em of a] {\(V_{k,k}\)};
     \vertex [above=1em of i2] {\(\langle a \rangle_f\)};
     \vertex [below=0.5em of b] {\(V_{k,k}\)};
    \diagram*{
    (a) -- [charged scalar, edge label'=\(0\)] (i1),
    (b) -- [charged scalar, edge label'=\(0\)] (i2),
    (b) -- [fermion, edge label=\(k\)] (a),
    (e2) -- [fermion, edge label=\(k\)] (b),
    (a) -- [fermion, edge label=\(k\)] (e1),
    };
  \end{feynman}
\end{tikzpicture}}}
+... , \label{G_av_diagram}
\end{align}
which obey simple Feynman rules. The solid-lined propagators with momentum $k$ denote a matrix-valued free GF $G^0(k, i \omega)$. The dashed propagators denote an also matrix-valued factor $V_{q_\mathrm{L},q_\mathrm{R}}$, where $q_\mathrm{L}$ is the momentum leaving the vertex of the dashed and the two solid propagators to the left and $q_\mathrm{R}$ the momentum joining it from the right. A vertex of $m$ dashed propagators obtains the $m$-th moment of the distribution $\langle a^m \rangle_f$ as a prefactor. The dashed propagators formally carry the momentum $q_\mathrm{R} - q_\mathrm{L}$ and all momenta joining a vertex of multiple dashed propagators add up to zero.  A sum $\frac{1}{N_\mathrm{sites}}\sum_q$ over all momenta inside a closed loop is implied.

Now the series can be rearranged by defining the self-energy $\Sigma(k, i\omega)$ as the sum of all irreducible diagrams that cannot be separated by cutting a single propagator

\begin{align}
G_{k,k}^{av}(i \omega)&=
\vcenter{\hbox{
\begin{tikzpicture}
  \begin{feynman}
     \vertex (e1) at (-0.5, -0.5);
     \vertex (e2) at (0.5, -0.5);
    \diagram*{
    (e2) -- [fermion, edge label=\(k\)] (e1),
    (e2) -- [fermion, half left, opacity = 0] (e1),
    (e1) -- [fermion, half left, opacity = 0] (e2),
    };
  \end{feynman}
\end{tikzpicture}}}
+
\vcenter{\hbox{
\begin{tikzpicture}
  \begin{feynman}
     \vertex (e1) at (-1, 0);
     \vertex (e2) at (2, 0);
     \vertex (a) at (0, 0);
     \vertex (b) at (1, 0);
     \vertex [right=0.7em of a] {\(\Sigma\)};
    \diagram*{
    (b) -- [fermion,half left] (a),
    (a) -- [fermion, half left] (b),
    (e2) -- [fermion, edge label=\(k\)] (b),
    (a) -- [fermion, edge label=\(k\)] (e1),
    };
  \end{feynman}
\end{tikzpicture}}}
+
\vcenter{\hbox{
\begin{tikzpicture}
  \begin{feynman}
     \vertex (e1) at (-1, 0);
     \vertex (e2) at (4, 0);
     \vertex (a) at (0, 0);
     \vertex (b) at (1, 0);
     \vertex (c) at (2, 0);
     \vertex (d) at (3, 0);
     \vertex [right=0.7em of a] {\(\Sigma\)};
     \vertex [right=0.7em of c] {\(\Sigma\)};
    \diagram*{
    (b) -- [fermion,half left] (a),
    (a) -- [fermion, half left] (b),
    (d) -- [fermion,half left] (c),
    (c) -- [fermion, half left] (d),
    (e2) -- [fermion, edge label=\(k\)] (d),
    (c) -- [fermion, edge label=\(k\)] (b),
    (a) -- [fermion, edge label=\(k\)] (e1),
    };
  \end{feynman}
\end{tikzpicture}}}
+... \nonumber \\
&=  G^0(k, i \omega) + G^0(k, i \omega) \Sigma(k, i\omega)\left[G^0(k, i \omega) + G^0(k, i \omega) \Sigma(k, i\omega) G^0(k, i \omega) + ...\right] \nonumber\\
&=  G^0(k, i \omega) + G^0(k, i \omega) \Sigma(k, i\omega)\left[G_{k,k}^{av}(i \omega)\right]. \nonumber
\end{align}
Because the free Green's function is given by $G^0(k, i \omega)=  [\mathbbm{1}i \omega- H_0(k)]^{-1}$, we finally obtain

\begin{align}
G_{k,k}^{av}(i \omega) = \left[\mathbbm{1} - \left(H_0(k) + \Sigma(k, i\omega)\right) \right]^{-1}. \nonumber 
\end{align}
As was said, the self-energy consists of all irreducible diagrams and assumes the form

\begin{align}
\Sigma(k, i\omega)=
\vcenter{\hbox{
\begin{tikzpicture}
  \begin{feynman}
     \vertex (a) at (0, 0);
     \vertex (i1) at (0, 4)[crossed dot]{};
     \vertex [above=1em of i1] {\(\langle a \rangle_f\)};
     \vertex [below=0.5em of a] {\(V_{k,k}\)};
    \diagram*{
    (a) -- [charged scalar, edge label'=\(0\)] (i1),
    };
  \end{feynman}
\end{tikzpicture}}}
+
\vcenter{\hbox{
\begin{tikzpicture}
  \begin{feynman}
     \vertex (a) at (0, 0);
     \vertex (i1) at (1.5, 4)[crossed dot]{};
     \vertex (b) at (3, 0);
     \vertex [above=1em of i1] {\(\langle a^2 \rangle_f\)};
     \vertex [below=0.5em of a] {\(V_{k,q}\)};
     \vertex [below=0.5em of b] {\(V_{q,k}\)};
    \diagram*{
      (b) -- [fermion, edge label=\(q\)] (a) ,
      (a) -- [charged scalar, edge label=\(q-k\)] (i1),
      (b) -- [charged scalar, edge label'=\(k-q\)] (i1),
    };
  \end{feynman}
\end{tikzpicture}}}
+
\vcenter{\hbox{
\begin{tikzpicture}
  \begin{feynman}
     \vertex (a) at (0, 0);
     \vertex (i1) at (1.5, 4)[crossed dot]{};
     \vertex (i2) at (1.5, 2)[crossed dot]{};
     \vertex (b) at (1.5, 0);
     \vertex (c) at (3, 0);
     \vertex [above=1em of i1] {\(\langle a^2 \rangle_f\)};
     \vertex [above=1em of i2] {\(\langle a \rangle_f\)};
     \vertex [below=0.5em of a] {\(V_{k,q}\)};
     \vertex [below=0.5em of b] {\(V_{q,q}\)};
     \vertex [below=0.5em of c] {\(V_{q,k}\)};
    \diagram*{
      (b) -- [fermion, edge label=\(q\)] (a) ,
      (a) -- [charged scalar, edge label=\(q-k\)] (i1),
      (b) -- [charged scalar, edge label'=\(0\)] (i2),
      (c) -- [charged scalar, edge label'=\(k-q\)] (i1),
      (c) -- [fermion, edge label=\(q\)] (b) ,
    };
  \end{feynman}
\end{tikzpicture}}}
+... \nonumber 
\end{align}
In FBA we only consider the first two diagrams of the series. The real-time frequency space GF and self-energy can be found through analytical continuation to the real axis, which simply amounts to replacing $i \omega \to \omega + i \eta$ with some infinitesimal regularization $\eta = 0^+$. \\

\section{An Analytical Expression for the AH Part of the Self-Energy}
\label{Analytical_AH_SE}
The disorder-induced self-energy correction in FBA assumes the form
\begin{align}
\Sigma(k, \omega + i \eta) &=  \langle a \rangle_f V_{k,k} + \frac{1}{N_\mathrm{site}} \langle a ^2 \rangle_f \sum_q V_{k,q}G^0( \omega 
+ i \eta, q) V_{q,k},  \nonumber \\
&=  \langle a \rangle_f V_{k,k} + \langle a ^2 \rangle_f  \frac{1}{2\pi} \int_{-\pi}^{\pi}\mathrm{d}q V_{k,q} G^0(\omega + i \eta, q)V_{q,k}, \label{SE_FBA_supp}
\end{align}
where we have assumed the continuum limit $\frac{1}{N_\mathrm{sites}}\sum_q \rightarrow \frac{1}{2\pi} \int_{-\pi}^{\pi}\mathrm{d}q$ in the second line.
Since hermiticity requires that $V_{k,q}^\dagger = V_{q,k}$, we have $V_{k,k}^\dagger = V_{k,k}$ and
$(V_{k,q} G^0(\omega + i \eta, q)V_{q,k})^\dagger = V_{k,q} G^0(\omega + i \eta, q)^\dagger V_{q,k}$, so the self-energy can only obtain an AH part 
if the free retarded Green's function $G^0(\omega + i \eta, q)$ is non-Hermitian. Because of that, we will decompose $G^0$ into a Hermitian and an AH part as the first step.

 Labeling the adjugate matrix as adj[...] and the eigenenergies/bands of $H_0(q)$ as $\{E_j(q)\}$, the Green's function can be written as 

\begin{align}
G^0(\omega + i \eta, q) &= \left(\mathbbm{1}(\omega + i \eta) - H_0(q)\right)^{-1}\nonumber
=\frac{1}{\mathrm{det}[\mathbbm{1}(\omega + i \eta) - H_0(q)]} \mathrm{adj}[\mathbbm{1}(\omega + i \eta) - H_0(q)] \nonumber \\
&= \frac{1}{ \displaystyle \prod_{j =1}^n (\omega + i \eta - E_j(q)) } \mathrm{adj}[\mathbbm{1}(\omega + i \eta) 
- H_0(q)].\nonumber \\
&\stackrel{\eta \to 0}{=} \prod_{j =1}^n \left( \frac{1}{\omega - E_j(q)} - i \pi \delta(\omega - E_j(q))\right)  \mathrm{adj}[\mathbbm{1}\omega
- H_0(q)]. \label{G_0}
\end{align}
In the last line we used the Dirac identity $\lim_{\eta \to 0} \frac{1}{i \eta + f(x)} = \frac{1}{f(x)} - i \pi \delta(f(x))$. If we now consider values of $\omega$ such that two bands never cross it at the same momentum $q$, all products containing more than one $\delta$-function from Eq. (\ref{G_0}) vanish. We are left with

\begin{align}
G^0(\omega , q) &=\left (\underbrace{ \prod_{j =1}^n \left( \frac{1}{\omega - E_j(q)}\right)}_{1/\mathrm{det}[\mathbbm{1}\omega - H_0(q)]}
 - i \pi \sum_{m=1}^n  \delta(\omega - E_m(q)) \prod_{\substack{j =1\\ j \neq m}}^n \left( \frac{1}{\omega - E_j(q)} \right) \right ) 
\mathrm{adj}[\mathbbm{1}\omega - H_0(q)] \nonumber \\
&=  (\mathbbm{1}\omega  - H_0(k))^{-1} - i \pi \left(\sum_{m=1}^n  \delta(\omega - E_m(k)) \prod_{\substack{j =1\\ j \neq m}}^n \left( \frac{1}{\omega - E_j(k)} \right)\right) \mathrm{adj}[\mathbbm{1}\omega - H_0(q)]. \label{G0_final}
\end{align}

The free Green's function has been decomposed into a Hermitian and an AH part. By substituting $\delta(\omega - E_m(q)) = \sum_{k_m \in K_m} \frac{\delta(q - k_m(\omega))}{\lvert E_m'(k_m(\omega)) \rvert }$, with $K_m$ denoting the set of momenta at which the band $E_m(q)$ intersects $\omega$, and plucking Eq. (\ref{G0_final}) into Eq. (\ref{SE_FBA_supp}), the AH self-energy contribution is readily obtained as 

\begin{align}
&\Sigma^{\mathrm{AH}}(k, \omega)  =  -\frac{i}{2} \langle a ^2 \rangle_f  \sum_{m=1}^{n}\sum_{k_m \in K_m}  \prod_{\substack{j =1\\ j \neq m}}^n  
\frac{1}{\omega - E_j(k_m(\omega))} 
 \frac{1}{\lvert E_m'(k_m(\omega)) \rvert }   V_{k,k_m(\omega)}\mathrm{adj}[\mathbbm{1}\omega - H_0(k_m(\omega))]V_{k_m(\omega),k}. \label{se_ah_supp}
\end{align}

\section{Application to a Single Nodal Point and Projection Onto the Subspace}
\label{Analytical_AH_SE}
\begin{wrapfigure}{r}[0cm]{8cm}  
        \vspace{- 20pt}
        \centering
        \includegraphics[width=8cm]{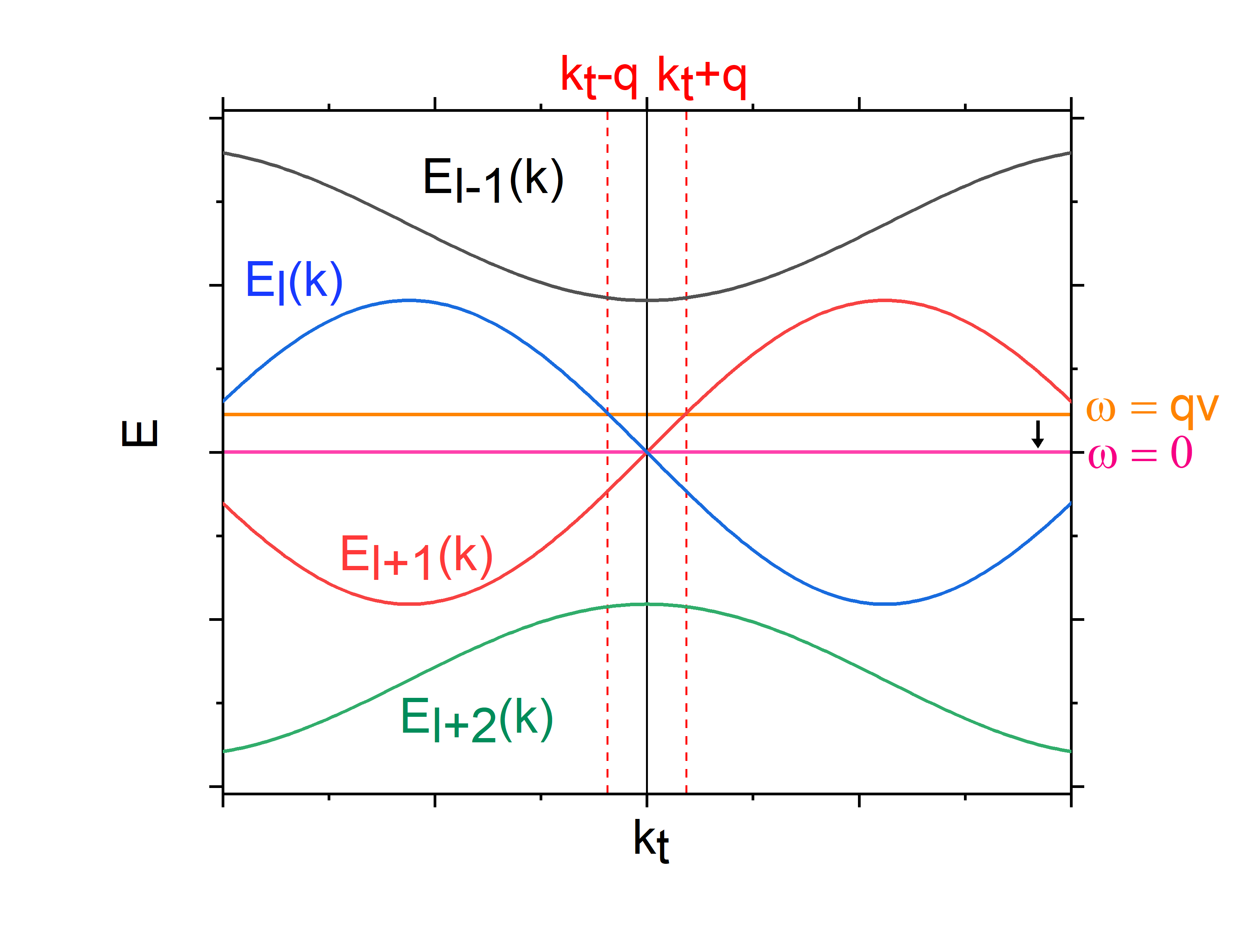}
        \caption{Linearization around $k_\mathrm{t}$.}  
        \label{Sfig1}      
\end{wrapfigure}

Now we will apply the findings from the previous section to a model with a single nodal point at $\omega = 0$. We assume that two bands $E_l$ and $E_{l+1}$ cross with a 
slope $v = \lvert E'_l(k_\mathrm{t})\rvert = \lvert E'_{l+1}(k_\mathrm{t})\rvert$ at some momentum $k_\mathrm{t}$. As we are interested in an effective Hamiltonian or respectively the self-energy correction at $\omega = 0$,  we take the limit $\omega \to 0$ of Eq. (\ref{se_ah_supp}) by linearizing $E_l(k_\mathrm{t} + q)  = -v q$, $E_{l +1}(k_\mathrm{t} + q)  =  v q$ and calculating $\Sigma^{\mathrm{AH}}(k, \omega =  0 ) = \lim_{q \to 0} \Sigma^{\mathrm{AH}}(k, \omega =  vq )$. The bands intersect $\omega = qv$ at $k_\mathrm{t} \pm q$, which is schematically depicted in Fig. \ref{Sfig1}. \\
\vspace{50pt}

Under these assumptions, Eq. (\ref{se_ah_supp}) tells us that

\begin{align}
\Sigma^{\mathrm{AH}}(k, q v)  = & -\frac{i}{2}  \langle a ^2 \rangle_f 
\left(\prod_{\substack{j =1\\ j \neq l, l+1}}^n  \frac{1}{ q v - E_j(k_\mathrm{t} - q)} \right)
\frac{1}{ q v - E_{l+1}(k_\mathrm{t} - q)} \frac{1}{\lvert E_l'(k_\mathrm{t} - q) \rvert} 
 V_{k,(k_\mathrm{t} - q)}\mathrm{adj}[\mathbbm{1}q v - H_0(k_\mathrm{t} - q)]V_{(k_\mathrm{t} - q),k} \nonumber \\
&-\frac{i}{2} \langle a ^2 \rangle_f 
\left(\prod_{\substack{j =1\\ j \neq l, l+1}}^n  \frac{1}{ q v - E_j(k_\mathrm{t} + q)} \right)
\frac{1}{q v - E_{l}(k_\mathrm{t} + q)}  \frac{1}{\lvert E_{l+1}'(k_\mathrm{t} + q) \rvert} 
 V_{k,(k_\mathrm{t} + q)}\mathrm{adj}[\mathbbm{1}qv - H_0(k_\mathrm{t} + q)]V_{(k_\mathrm{t} + q),k} \nonumber  \\
= & -\frac{i}{2}  \langle a ^2 \rangle_f 
\left(\prod_{\substack{j =1\\ j \neq l, l+1}}^n  \frac{1}{ q v - E_j(k_\mathrm{t} - q)} \right)
\frac{1}{ 2 q v } \frac{1}{v} 
 V_{k,(k_\mathrm{t} - q)}\mathrm{adj}[\mathbbm{1}q v - H_0(k_\mathrm{t} - q)]V_{(k_\mathrm{t} - q),k} \nonumber \\
&-\frac{i}{2} \langle a ^2 \rangle_f  \left(\prod_{\substack{j =1\\ j \neq l, l+1}}^n  \frac{1}{ q v - E_j(k_\mathrm{t} + q)} \right)
\frac{1}{2q v }  \frac{1}{v} 
 V_{k,(k_\mathrm{t} + q)}\mathrm{adj}[\mathbbm{1}qv - H_0(k_\mathrm{t} + q)]V_{(k_\mathrm{t} + q),k}. \nonumber  
\end{align}
From this, we see  that
\begin{align}
\Sigma^{\mathrm{AH}}(k, 0 ) =& \lim_{q \to 0} \Sigma^{\mathrm{AH}}(k,  vq ) \nonumber \\ 
=& -\frac{i}{2v} \langle a ^2 \rangle_f \left(\prod_{\substack{j =1\\ j \neq l, l+1}}^n  \frac{1}{ - E_j(k_\mathrm{t})}\right)
V_{k,k_\mathrm{t}} \lim_{q \to 0}\biggl(\frac{1}{2qv}
\mathrm{adj}[\mathbbm{1}qv- \mathcal{H}_0(k_\mathrm{t} - q)])\nonumber \\
&+ \frac{1}{2qv} \mathrm{adj}[\mathbbm{1}qv - \mathcal{H}_0(k_\mathrm{t} + q)] \biggr) V_{k_\mathrm{t},k}. \label{se_ah_dirac}
\end{align}
Next, we use the unitary transformation $U_k$ to the eigenbasis of $H_0(k)$ and insert an identity of the form 
$U_{(k_\mathrm{t} - q)}^\dagger U_{(k_\mathrm{t} - q)}$ into the first adjugate from Eq. (\ref{se_ah_dirac}). Further we use that
$\mathrm{adj}[AB] = \mathrm{adj}[B] \mathrm{adj}[A]$ and obtain

\begin{align}
\frac{1}{2qv}\mathrm{adj}[\mathbbm{1}qv - H_0(k_\mathrm{t} - q)]  =&\frac{1}{2qv} \mathrm{adj}[U_{(k_\mathrm{t} - q)}^\dagger U_{(k_\mathrm{t} - q)} (\mathbbm{1}qv - H_0(k_\mathrm{t} - q))U_{(k_\mathrm{t} - q)}^\dagger U_{(k_\mathrm{t} - q)}] \nonumber \\
=&\frac{1}{2qv}\mathrm{adj}[U_{(k_\mathrm{t} - q)}]  \mathrm{adj}[\mathbbm{1}qv
- U_{(k_\mathrm{t} - q)}  H_0(k_\mathrm{t} - q) U_{(k_\mathrm{t} - q)}^\dagger] 
 \mathrm{adj}[U_{(k_\mathrm{t} - q)}^\dagger ]. \nonumber
\end{align}
For small $q$ it follows trivially that
\begin{align}
U_{(k_\mathrm{t} - q)}  H_0(k_\mathrm{t} - q) U_{(k_\mathrm{t} - q)}^\dagger = 
\begin{bmatrix}
E_1(k_\mathrm{t} - q)  &               &                                           &                                          &               &  \\
                       & \ddots        &                                           &                                          & \text{\large 0}&  \\
                       &               &  \tikzmark{left1}qv & 0                                        &               &  \\
                       &               &  0                                        & -qv \tikzmark{right1}&               &  \\
                       & \text{\large 0}&                                           &                                          & \ddots        &  \\
                       &               &                                           &                                          &               &  E_n(k_\mathrm{t} - q)
\end{bmatrix}, \nonumber
\DrawBox[thick, red ]{left1}{right1}{}
\end{align}
where the red box encloses the matrix elements belonging to the subspace spanned by the two touching bands $E_l$ and $E_{l+1}$. Thusly,  
$\frac{1}{2qv}\mathrm{adj}[\mathbbm{1}qv - H_0(k_\mathrm{t} - q)]$ takes the form

\begin{align}
&\frac{1}{2qv} \mathrm{adj}[U_{(k_\mathrm{t} - q)}]  \mathrm{adj}\begin{bmatrix}
(q v - E_1(k_\mathrm{t} - q))  &               &                       &                      &               &  \\
                                                   & \ddots        &                       &                      & \text{\large 0}&  \\
                                                   &               & \tikzmark{left2} 0    & 0                    &               &  \\
                                                   &               & 0                     & 2qv \tikzmark{right2}&               &  \\
                                                   & \text{\large 0}&                       &                      & \ddots        &  \\
                                                   &               &                       &                      &               &  ( q v - E_n(k_\mathrm{t} - q))
\end{bmatrix}   \mathrm{adj}[U_{(k_\mathrm{t} - q)}^\dagger ] \nonumber \\
 \nonumber \\
=& \frac{1}{2qv}\mathrm{adj}[U_{(k_\mathrm{t} - q)}]  \left(\prod_{\substack{j =1\\ j \neq l, l+1}}^n  ( q v - E_j(k_\mathrm{t} - q))\right)	
\begin{bmatrix}
0  &               &                       &                      &               &  \\
   & \ddots        &                       &                      & \text{\large 0}&  \\
   &               & \tikzmark{left3} 0    & 0                    &               &  \\
   &               & 0                     & 2qv \tikzmark{right3}&               &  \\
   & \text{\large 0}&                       &                      & \ddots        &  \\
   &               &                       &                      &               &  0
\end{bmatrix} \mathrm{adj}[U_{(k_\mathrm{t} - q)}^\dagger ] \nonumber \\
=&\mathrm{adj}[U_{(k_\mathrm{t} - q)}]  \left(\prod_{\substack{j =1\\ j \neq l, l+1}}^n  ( q v - E_j(k_\mathrm{t} - q))\right)	
\begin{bmatrix}
0  &               &                       &                      &               &  \\
   & \ddots        &                       &                      & \text{\large 0}&  \\
   &               & \tikzmark{left4} 0    & 0                    &               &  \\
   &               & 0                     & 1 \tikzmark{right4}  &               &  \\
   & \text{\large 0}&                       &                      & \ddots        &  \\
   &               &                       &                      &               &  0
\end{bmatrix}  \mathrm{adj}[U_{(k_\mathrm{t} - q)}^\dagger ] \nonumber.
\DrawBox[thick, red ]{left2}{right2}{}
\DrawBox[thick, red ]{left3}{right3}{}
\DrawBox[thick, red ]{left4}{right4}{}
\end{align}
The second line becomes obvious if we remember the element-wise definition of the adjugate matrix as $(\mathrm{adj}[A])_{i,j} = (-1)^{i+j} M_{j,i}$, where the minor $M_{j,i}$ of $A$ denotes the determinant of the matrix that is obtained by deleting row $j$ and column $i$ of A.\\

The second term $\frac{1}{2qv} \mathrm{adj}[\mathbbm{1}qv - H_0(k_\mathrm{t} + q)]$ in Eq. (\ref{se_ah_dirac}) can be treated in the same fashion to read

\begin{align}
=&\mathrm{adj}[U_{(k_\mathrm{t} + q)}]  \left(\prod_{\substack{j =1\\ j \neq l, l+1}}^n  ( q v - E_j(k_\mathrm{t} + q))\right)	
\begin{bmatrix}
0  &               &                       &                      &               &  \\
   & \ddots        &                       &                      & \text{\large 0}&  \\
   &               & \tikzmark{left4} 1    & 0                    &               &  \\
   &               & 0                     & 0 \tikzmark{right4}  &               &  \\
   & \text{\large 0}&                       &                      & \ddots        &  \\
   &               &                       &                      &               &  0
\end{bmatrix}  \mathrm{adj}[U_{(k_\mathrm{t} + q)}^\dagger ] \nonumber.
\DrawBox[thick, red ]{left2}{right2}{}
\DrawBox[thick, red ]{left3}{right3}{}
\DrawBox[thick, red ]{left4}{right4}{}
\end{align}
With these results, Eq. (\ref{se_ah_dirac}) becomes

\begin{align}
\Sigma^{\mathrm{AH}}(k, 0 ) &= -\frac{i}{2v}\langle a ^2 \rangle_f  V_{k,k_\mathrm{t}}  \mathrm{adj}[U_{k_\mathrm{t}}]
\begin{bmatrix}
0  &               &                       &                      &               &  \\
   & \ddots        &                       &                      & \text{\large 0}&  \\
   &               & \tikzmark{left6} 1    & 0                    &               &  \\
   &               & 0                     & 1 \tikzmark{right6}  &               &  \\
   & \text{\large 0}&                       &                      & \ddots        &  \\
   &               &                       &                      &               &  0
\end{bmatrix}   \mathrm{adj}[U_{k_\mathrm{t}}^\dagger]V_{k_\mathrm{t},k}. \nonumber\\
&= -\frac{i}{2v}\langle a ^2 \rangle_f  V_{k,k_\mathrm{t}}  U_{k_\mathrm{t}}^\dagger
\begin{bmatrix}
0  &               &                       &                      &               &  \\
   & \ddots        &                       &                      & \text{\large 0}&  \\
   &               & \tikzmark{left7} 1    & 0                    &               &  \\
   &               & 0                     & 1 \tikzmark{right7}  &               &  \\
   & \text{\large 0}&                       &                      & \ddots        &  \\
   &               &                       &                      &               &  0
\end{bmatrix}   U_{k_\mathrm{t}}V_{k_\mathrm{t},k}, \label{se_ah_dirac_2}
\DrawBox[thick, red ]{left6}{right6}{}
\DrawBox[thick, red ]{left7}{right7}{}
\end{align}
which can be used to calculate the projection $\widetilde{\Sigma}^{\mathrm{AH}}(k_\mathrm{t}, 0) $ of the AH self-energy contribution onto the subspace of the crossing bands at $k_\mathrm{t}$. Let $\widetilde{V}_{k_\mathrm{t}, k_\mathrm{t}}$ denote the subspace-projection of $V_{k_\mathrm{t}, k_\mathrm{t}}$ such that

\begin{align}
U_{k_\mathrm{t}}V_{k_\mathrm{t},k_\mathrm{t}} U_{k_\mathrm{t}}^\dagger =
\begin{bmatrix}
   & \ddots &                     &                       & \ddots \\
   &        & \tikzmark{left8} ...& ...                   &        \\
   &        & ...                 & ...\tikzmark{right8}  &        \\
   & \ddots &                     &                       & \ddots
\end{bmatrix} .\nonumber
\DrawBox[thick, red ]{left8}{right8}{$\widetilde{V}_{k_\mathrm{t}, k_\mathrm{t}}$}
\end{align}
Then Eq. (\ref{se_ah_dirac_2}) yields
\begin{align}
U_{k_\mathrm{t}} \Sigma^{\mathrm{AH}}(k_\mathrm{t}, 0 ) U_{k_\mathrm{t}}^\dagger 
=& -\frac{i}{2v} \langle a ^2 \rangle_f U_{k_\mathrm{t}} V_{k_\mathrm{t},k_\mathrm{t}} U_{k_\mathrm{t}}^\dagger 
\begin{bmatrix}
0  &               &                       &                      &               &  \\
   & \ddots        &                       &                      & \text{\large 0}&  \\
   &               & \tikzmark{left9} 1    & 0                    &               &  \\
   &               & 0                     & 1 \tikzmark{right9}  &               &  \\
   & \text{\large 0}&                       &                      & \ddots        &  \\
   &               &                       &                      &               &  0
\end{bmatrix} U_{k_\mathrm{t}}V_{k_\mathrm{t},k_\mathrm{t}}U_{k_\mathrm{t}}^\dagger \nonumber \\
=& -\frac{i}{2v} \langle a ^2 \rangle_f
\begin{bmatrix}
   & \ddots &                     &                       & \ddots \\
   &        & \tikzmark{left10} ...& ...                   &        \\
   &        & ...                 & ...\tikzmark{right10}  &        \\
   & \ddots &                     &                       & \ddots
\end{bmatrix} ,\nonumber
\DrawBox[thick, red ]{left9}{right9}{}
\DrawBox[thick, red ]{left10}{right10}{$\widetilde{V}_{k_\mathrm{t}, k_\mathrm{t}}^2$}
\end{align}
and thus we arrive at the expression used in the main text 
\begin{align}
\widetilde{\Sigma}^{\mathrm{AH}}(k_\mathrm{t},  0) =-\frac{i}{2v} \langle a ^2 \rangle_f (\widetilde{V}_{k_\mathrm{t}, k_\mathrm{t}})^2. \label{Sigma_AH_Projection_Supp}
\end{align}

Lastly in this section, we elaborate on the projection of the chiral symmetry onto the subspace. The chiral symmetry in general indicates that there exists some matrix $\Gamma$ acting on the internal degrees of freedom such that $\mathbbm{1} \otimes \Gamma (H_0 + V) \mathbbm{1} \otimes \Gamma^{-1} = -(H_0 + V)$, where $\mathbbm{1}$ is the $N_\mathrm{site} \times N_\mathrm{site}$ identity matrix. This implies that $ \Gamma H_0(k) \Gamma^{-1} = -H_0(k)$ and $ \Gamma V_{k, k} \Gamma^{-1} = -V_{k, k}$. \\

Due to the chiral symmetry, the eigenvalues  come in positive and negative pairs, which we denote here by $E^+_j(k)$ and $E^-_j(k) = -E^+_j(k)$.  We can transform $H_0(k)$ into its eigenbasis, where we sort the eigenvectors in pairs of positive and negative eigenvalue. The transformed Hamiltonian is denoted by a hat and reads 

\begin{align}
\hat{H}_0(k) &= \mathrm{diag}[E^+_1(k), E^-_1(k), ..., E^+_l(k), E^-_l(k),..., E^+_n(k), E^-_n(k)] \nonumber 
\end{align}
The pair $E^+_l$, $E^-_l$ is again the one that touches at zero energy. Next, we consider the transformation of $\Gamma$ into this basis, denoted by $\hat{\Gamma}(k)$ .  Since we only performed a unitary basis transformation, $\hat{H}_0(k) \hat{\Gamma}(k) = -\hat{\Gamma}(k) \hat{H}_0(k)$ still holds and thus 
$\hat{H}_0(k) \hat{\Gamma}(k) \lvert E^+_j(k)\rangle = -\hat{\Gamma}(k) \hat{H}_0(k) \lvert E^+_j(k)\rangle = - E^+_j(k)\hat{\Gamma}(k)  \lvert E^+_j(k)\rangle$ for any $j = 1,...,n$. 
Apparently, $\hat{\Gamma}(k) \lvert E^+_j(k)\rangle$ is an eigenstate of  $\hat{H}_0(k)$ with eigenvalue $ - E^+_j(k)$. As long as the eigenstates are non-degenerate, there is only one such eigenstate, so $\hat{\Gamma}(k) \lvert E^+_j(k)\rangle = \lambda^-_j \lvert E^-_j(k)\rangle$ with some $\lambda^-_j \in  \mathbbm{C}$ and similarly $\hat{\Gamma}(k) \lvert E^-_j(k)\rangle = \lambda^+_j \lvert E^+_j(k)\rangle$ with some $\lambda^+_j \in  \mathbbm{C}$. In our basis, the eigenstates of $\hat{H}_0(k)$ are simply unit vectors and the $\pm$-pairs are situated on top of each other, for example $\lvert E^+_1(k)\rangle = (1,0, 0,...,0)^\mathrm{T}$, $\lvert E^-_1(k)\rangle = (0,1,0,...,0)^\mathrm{T}$ and so on. Thereby, $\hat{\Gamma}(k)$ must be $2\times2$-block-diagonal, with blocks 
$\gamma_{j,x}(k) \sigma_x + \gamma_{j,y}(k) \sigma_y$ for each energy pair. We write

\begin{align}
\hat{\Gamma}(k) = \mathrm{diag}[\gamma_{1,x}(k) \sigma_x + \gamma_{1,y}(k) \sigma_y, ..., \underbrace{\gamma_{l,x}(k) \sigma_x + \gamma_{l,y}(k) \sigma_y}_{\widetilde{\Gamma}(k)},..., 
\gamma_{n,x}(k) \sigma_x + \gamma_{n,y}(k) \sigma_y]. \label{Gamma} 
\end{align}
The block belonging to the $l$-th energy pair (the one that touches at zero energy) is exactly the projection of $\Gamma$ onto the subspace, which we consistently denote by $\widetilde{\Gamma}(k)$.
Up to now, we only considered values of $k$ with no degeneracy in the spectrum. However, for a physically sensible Hamiltonian $H_0(k)$ degeneracies should only appear at isolated points in the parameter space, so $\hat{\Gamma}(k)$ cannot deviate from the block form of Eq. (\ref{Gamma}) at these points due to continuity. Of course, $\hat{\Gamma}^{-1}(k)$ is also block-diagonal.\\

Now we can show that $\widetilde{\Gamma}(k) \widetilde{V}_{k, k} (\widetilde{\Gamma}(k))^{-1} = - \widetilde{V}_{k, k}$ for any $k$. If we transform the scattering vertex $V_{k,k}$ into the eigenbasis of $H_0(k)$, 
the chiral symmetry remains and $\hat{\Gamma}(k)\hat{V}_{k,k} \hat{\Gamma}^{-1}(k) = - \hat{V}_{k,k}$. We write this in terms of matrices and mark the blocks belonging to the subspace of 
$\lvert E^\pm_l(k)\rangle$ by a red box.

\begin{align}
\hat{\Gamma}(k)\hat{V}_{k,k} \hat{\Gamma}^{-1}(k) &=
\begin{bmatrix}
   & \ddots &                     &                       & \text{\large 0} \\
   &        & \tikzmark{left1} ...& ...                   &        \\
   &        & ...                 & ...\tikzmark{right1}  &        \\
   & \text{\large 0} &                     &                       & \ddots
\end{bmatrix} 
\begin{bmatrix}
   & \ddots &                     &                       & \ddots \\
   &        & \tikzmark{left2} ...& ...                   &        \\
   &        & ...                 & ...\tikzmark{right2}  &        \\
   & \ddots  &                     &                       & \ddots
\end{bmatrix} 
\begin{bmatrix}
   & \ddots &                     &                       & \text{\large 0} \\
   &        & \tikzmark{left3} ...& ...                   &        \\
   &        & ...                 & ...\tikzmark{right3}  &        \\
   & \text{\large 0} &                     &                       & \ddots
\end{bmatrix} 
= \begin{bmatrix}
   & \ddots &                     &                       & \ddots \\
   &        & \tikzmark{left4} ...& ...                   &        \\
   &        & ...                 & ...\tikzmark{right4}  &        \\
   & \ddots  \hspace{25pt}&                     &                       & \hspace{25pt} \ddots
\end{bmatrix} \nonumber \\
&= - \hat{V}_{k,k} = -\begin{bmatrix}
   & \ddots &                     &                       &  \ddots \\
   &        & \tikzmark{left5} ...& ...                   &        \\
   &        & ...                 & ...\tikzmark{right5}  &        \\
   &  \ddots &                     &                       & \ddots
\end{bmatrix} . \nonumber
\DrawBox[thick, red ]{left1}{right1}{$\widetilde{\Gamma}(k)$}
\DrawBox[thick, red ]{left2}{right2}{$\widetilde{V}_{k, k}$}
\DrawBox[thick, red ]{left3}{right3}{$(\widetilde{\Gamma}(k))^{-1}$}
\DrawBox[thick, red ]{left4}{right4}{$\widetilde{\Gamma}(k) \widetilde{V}_{k, k} (\widetilde{\Gamma}(k))^{-1}$}
\DrawBox[thick, red ]{left5}{right5}{$\widetilde{V}_{k, k}$}
\end{align}
Comparing the first and second line shows that indeed
\begin{align}
\widetilde{\Gamma}(k) \widetilde{V}_{k, k} (\widetilde{\Gamma}(k))^{-1} = - \widetilde{V}_{k, k}. \nonumber
\end{align}

\section{Scattering Between two Nodal Points and Necessity of the Correlation Between OS- and NN-Terms in the Random Disorder}
\label{nodal_point_spliting}
Consider a general model $H = H_0 + V$ with two internal degrees of freedom and a chiral symmetry such that $\sigma_z H \sigma_z = - H$, similar to the model studied in the main text. This implies the symmetry 
$\sigma_z H_e(k, w)^\dagger \sigma_z = - H_e(k, -w)$ for the effective Hamiltonian $ H_e(k, w) = H_0(k) +  \Sigma(k, w)$ emerging from the averaged GF description, 
so $\sigma_z H^\dagger_e(k, 0) \sigma_z = - H_e(k, 0)$ and thus $d_x, d_y \in \mathbb{R}$ and $d_0, d_z \in i\mathbb{R}$ 
if we parametrize $H_e(k, 0) = d_0 \sigma_0 + \boldsymbol{d} \cdot \boldsymbol{\sigma}$. \\

The spectrum of a Matrix $d_0 \sigma_0 + (\bm d_R +  i \bm d_I) \bm \sigma$ with $\bm d_R, \bm d_I \in \mathbb{R}$ is given 
by $E_\pm=d_0 \pm \sqrt{\bm d_R^2 - \bm d_I^2 + 2 i\bm d_R \cdot \bm d_I}$ and exceptional points occur if $\bm d_R^2 - \bm d_I^2 = 0$ and $\bm d_R \cdot  \bm d_I= 0$ is satisfied simultaneously while $\bm d_R, \bm d_I \neq 0$. In conclusion, an $i \sigma_z$ contribution in the self-energy $\Sigma(k, 0)$ is required to open an exceptional point in the spectrum of $H_e(k, 0)$ in our chirally symmetric model. Using Eq. (\ref{Sigma_AH_Projection_Supp}), we already discussed in the main text that such a term is impossible directly at the band touching point $k_\mathrm{t}$ of a model with only a single nodal point at zero energy. However, if $H_0(k)$ exhibits two nodal points at 
momenta $k_\mathrm{t_1}$ and $k_\mathrm{t_2}$, where the bands cross with slopes $v_1$ and $v_2$, a ``scattering'' process between the two points can create 
an $i \sigma_z$ term, as we will show in the following.\\

Due to the symmetry constraints, the impurity matrix elements from Eq. (\ref{impurity}) take the form $V_{k,k'} = \alpha_{k,k'}\sigma_x + \beta_{k,k'}\sigma_y$. We can apply Eq. (\ref{se_ah_dirac_2}) to both points separately and directly read off the result for the AH part of the self-energy  in FBA

\begin{align}
\Sigma^{\mathrm{AH}}(k, 0 )  = & -\frac{i}{2v_1} \langle a ^2 \rangle_f 
V_{k,k_\mathrm{t_1}} V_{k_\mathrm{t_1},k} -\frac{i}{2v_2} \langle a ^2 \rangle_f  V_{k,k_\mathrm{t_2}} V_{k_\mathrm{t_2},k} \nonumber \\
=&\left(-\frac{i}{2v_1} \langle a ^2 \rangle_f (\lvert \alpha_{k,k_\mathrm{t_1}} \rvert ^2 
+ \lvert \beta_{k,k_\mathrm{t_1}} \rvert ^2) -\frac{i}{2v_2} \langle a ^2 \rangle_f ( \lvert \alpha_{k,k_\mathrm{t_2}} \rvert ^2 
+ \lvert \beta_{k,k_\mathrm{t_2}} \rvert ^2 ) \right) \sigma_0 \nonumber \\
& \left( \frac{i}{v_1} \langle a ^2 \rangle_f (\operatorname{Im}[\alpha_{k,k_\mathrm{t_1}}\beta_{k_\mathrm{t_1},k}]
+\frac{i}{v_2} \langle a ^2 \rangle_f   \operatorname{Im}[\alpha_{k,k_\mathrm{t_2}}\beta_{k_\mathrm{t_2},k}] \right) \sigma_z. \nonumber
\end{align} 
Due to hermiticity,  $V_{k,k'} ^\dagger = V_{k',k}$ must always hold true and thus the $i \sigma_z$ term created by the nodal point at $k_\mathrm{t_1}$ vanishes in the self-energy $\Sigma^{\mathrm{AH}}(k_\mathrm{t_1}, 0 )$ at $k_\mathrm{t_1}$, 
since $\operatorname{Im}[\alpha_{k_\mathrm{t_1},k_\mathrm{t_1}}\beta_{k_\mathrm{t_1},k_\mathrm{t_1}}] = 0$. This agrees fully with the arguments presented this far. On the other hand, the $i \sigma_z$ term created by the second nodal 
point $k_\mathrm{t_2}$ can persist. If we assume OS and NN impurities such that

\begin{align}
V_{k,k'} &=\left(V_\mathrm{OS_x} +  \left(V_\mathrm{NN_x}e^{-ik } + V^*_\mathrm{NN_x} e^{ik' }\right)\right) \sigma_x
+ \left(V_\mathrm{OS_y} + \left(V_\mathrm{NN_y}e^{-ik } + V^*_\mathrm{NN_y}e^{ik'}\right) \right) \sigma_y \nonumber \\
&= \left(V_\mathrm{OS_x} +  |V_\mathrm{NN_x}|\left(e^{-i(k + \Delta_x)} + e^{i(k' + \Delta_x)} \right)\right) \sigma_x
+ \left(V_\mathrm{OS_y} +  |V_\mathrm{NN_y}| \left(e^{-i(k + \Delta_y)} + e^{i(k' + \Delta_y)} \right) \right) \sigma_y, \nonumber
\end{align}
we get
$\operatorname{Im}[\alpha_{k_\mathrm{t_1},k_\mathrm{t_2}}\beta_{k_\mathrm{t_2},k_\mathrm{t_1}}] 
= V_\mathrm{OS_x} |V_\mathrm{NN_y}|(\sin{(k_\mathrm{t_1} + \Delta_x)} - \sin{(k_\mathrm{t_2} + \Delta_x)})
+   V_\mathrm{NN_x}V_\mathrm{OS_y}(\sin{(k_\mathrm{t_2} + \Delta_x)} - \sin{(k_\mathrm{t_1} + \Delta_x)})$, which is non-zero in general. The contribution to the AH self-energy from the second nodal point can open an EP at the first nodal point in $k$-space and vice versa, which is exactly what happens in our model from the main text.\\

Finally, we discuss why the correlation in amplitude between the OS and NN terms is necessary. Suppose that there were two different kinds of 
impurities with independent random amplitudes $\{a^\mathrm{I}_j \}$ and $\{a^\mathrm{II}_j \}$, such that $V$ from Eq. (\ref{impurity}) takes the form
\begin{align}
V = \sum_{j= 1} ^ {N_\mathrm{site}} \sum_{k,k'} \frac{e^{-i j (k- k')}}{N_\mathrm{sites}} c_k^\dagger \left (a^\mathrm{I}_j V^\mathrm{I}_{k,k'}
 + a^\mathrm{II}_j V^\mathrm{II}_{k,k'}\right) c_{k'}. \nonumber
\end{align}
The perturbation theory can be generalized easily to this case. The perturbation series for the full GF remains the same as in Eq. (\ref{PTB_series}) and we average the blocks  $G_{k,k'}(i \omega)$  over the distributions $f^\mathrm{I}$ and $f^\mathrm{II}$ of the amplitudes similarly to Eq. (\ref{G_av})

\begin{align}
G_{k,k'}^{av}(i \omega) &= \langle G_{k,k'}(i \omega) \rangle_{f^\mathrm{I},f^\mathrm{II}} \nonumber \\
&= \int \mathrm{d}a^\mathrm{I}_1 \mathrm{d}a^\mathrm{II}_1 f^\mathrm{I}(a^\mathrm{I}_1)  f^\mathrm{II}(a^\mathrm{II}_1) 
\int \mathrm{d}a^\mathrm{I}_2 \mathrm{d}a^\mathrm{II}_2 f^\mathrm{I}(a^\mathrm{I}_2)  f^\mathrm{II}(a^\mathrm{II}_2)  ... 
\int \mathrm{d}a^\mathrm{I}_{N_\mathrm{site}} \mathrm{d}a^\mathrm{II}_{N_\mathrm{site}}  f^\mathrm{I}(a^\mathrm{I}_{N_\mathrm{site}}) f^\mathrm{II}(a^\mathrm{II}_{N_\mathrm{site}})
G_{k,k'}(i \omega). \nonumber
\end{align} 
This expression contains terms of the form $\langle \sum^{N_\mathrm{site}}_{j_1, ...j_m = 1} a^\mathrm{I}_{j_1}a^\mathrm{II}_{j_2}...a^\mathrm{II}_{j_{m}}e^{\sum^m_{l=1} q_l j_l}\rangle_{f^\mathrm{I},f^\mathrm{II}} $. Again, we treat these terms by grouping them into terms where all scattering vectors $q \in \mathbf{Q} = \{q_1, q_2, ..., q_m\}$ are connected to one, two, three and so on impurities. The variables $m_{\mathrm{I}}$, $m_{\mathrm{II}}$ denote the total number of impurities of type I, II (so $m_{\mathrm{I}} + m_{\mathrm{II}} = m$), and $m_{\mathrm{I}, r}$, $m_{\mathrm{II}, r}$ denote the number of impurities of type I, II to which the momenta from the subset $\mathbf{Q}_r \subset \mathbf{Q} $ are connected.  

\begin{align}
\langle \sum^{N_\mathrm{site}}_{j_1, ...j_m = 1} a^\mathrm{I}_{j_1}a^\mathrm{II}_{j_2}...a^\mathrm{II}_{j_{m}}e^{\sum^m_{l=1} q_l j_l}\rangle_{f^\mathrm{I},f^\mathrm{II}}   = &
\langle \sum^{N_\mathrm{site}}_{h_1=1} (a^\mathrm{I}_{h_1})^{m_\mathrm{I}} (a^\mathrm{II}_{h_1})^{m_\mathrm{II}}
e^{\sum_{q \in \mathbf{Q}} q h_1}\rangle_{f^\mathrm{I},f^\mathrm{II}} \nonumber \\
&+ \langle \sum_{\cup^2_{r = 1} \mathbf{Q}_r = \mathbf{Q}} \sum^{N_\mathrm{site}}_{h_1=1} 
\sum^{N_\mathrm{site}}_{\substack{h_2 = 1\\ h_2 \neq h_1}}
(a^\mathrm{I}_{h_1})^{m_{\mathrm{I},1}} (a^\mathrm{II}_{h_1})^{m_{\mathrm{II},1}}
(a^\mathrm{I}_{h_2})^{m_{\mathrm{I},2}} (a^\mathrm{II}_{h_2})^{m_{\mathrm{II},2}} e^{\sum_{q_1 \in \mathbf{Q_1}} q_1 h_1} 
e^{\sum_{q_2 \in \mathbf{Q_2}} q_2 h_2}\rangle_f \nonumber \\
&+ \langle \sum_{\cup^3_{r = 1} \mathbf{Q}_r = \mathbf{Q}} \sum^{N_\mathrm{site}}_{h_1 = 1} 
\sum^{N_\mathrm{site}}_{\substack{h_2 = 1\\ h_2 \neq h_1}} \sum^{N_\mathrm{site}}_{\substack{h_3 = 1\\ h_3 \neq h_1, h_2}} 
(a^\mathrm{I}_{h_1})^{m_{\mathrm{I},1}} (a^\mathrm{II}_{h_1})^{m_{\mathrm{II},1}}
(a^\mathrm{I}_{h_2})^{m_{\mathrm{I},2}} (a^\mathrm{II}_{h_2})^{m_{\mathrm{II},2}}
(a^\mathrm{I}_{h_3})^{m_{\mathrm{I},3}} (a^\mathrm{II}_{h_3})^{m_{\mathrm{II},3}}    \nonumber \\
& \times e^{\sum_{q_1 \in \mathbf{Q_1}} q_1 h_1}
e^{\sum_{q_2 \in \mathbf{Q_2}} q_2 h_2} e^{\sum_{q_3 \in \mathbf{Q_3}} q_3 h_3}\rangle_f \nonumber\\
&+ ... \nonumber
\end{align}	
We introduce a small error of the order $\frac{1}{N_\mathrm{site}}$ by letting the $h$ 
sums run unrestricted, e.g. $\sum^{N_\mathrm{site}}_{\substack{h_2 = 1\\ h_2 \neq h_1}} \to \sum^{N_\mathrm{site}}_{h_2 = 1}$ and then take the average 
to obtain

\begin{align}
\langle \sum^{N_\mathrm{site}}_{j_1, ...j_m = 1} a^\mathrm{I}_{j_1}a^\mathrm{II}_{j_2}...a^\mathrm{II}_{j_{m}}e^{\sum^m_{l=1} q_l j_l}\rangle_{f^\mathrm{I},f^\mathrm{II}}  = &
N_\mathrm{site} \langle (a^\mathrm{I})^{m_\mathrm{I}} \rangle_{f^\mathrm{I}}  \langle (a^\mathrm{II})^{m_\mathrm{II}} \rangle_{f^\mathrm{II}}\delta_{0,\sum_{q \in \mathbf{Q}}q} \nonumber \\
&+ (N_\mathrm{site})^2\sum_{\cup^2_{r = 1} \mathbf{Q}_r = \mathbf{Q} }
\langle(a^\mathrm{I})^{m_{\mathrm{I},1}}\rangle_{f^\mathrm{I}} \langle(a^\mathrm{II})^{m_{\mathrm{II},1}}\rangle_{f^\mathrm{II}} 
\delta_{0,\sum_{q_1 \in \mathbf{Q_1}}q_1}\nonumber \\
& \times \langle(a^\mathrm{I})^{m_{\mathrm{I},2}} \rangle_{f^\mathrm{I}}\langle (a^\mathrm{II})^{m_{\mathrm{II},2}}\rangle_{f^\mathrm{II}} 
 \delta_{0,\sum_{q_2 \in \mathbf{Q_2}}q_2} \nonumber \\
&+ (N_\mathrm{site})^3\sum_{\cup^3_{r = 1} \mathbf{Q}_r = \mathbf{Q}} 
\langle(a^\mathrm{I})^{m_{\mathrm{I},1}}\rangle_{f^\mathrm{I}} \langle(a^\mathrm{II})^{m_{\mathrm{II},1}}\rangle_{f^\mathrm{II}}
\delta_{0,\sum_{q_1 \in \mathbf{Q_1}}q_1}
\langle(a^\mathrm{I})^{m_{\mathrm{I},2}} \rangle_{f^\mathrm{I}}\langle (a^\mathrm{II})^{m_{\mathrm{II},2}}\rangle_{f^\mathrm{II}}\nonumber \\
& \times\delta_{0,\sum_{q_2 \in \mathbf{Q_2}}q_2}
\langle(a^\mathrm{I})^{m_{\mathrm{I},3}} \rangle_{f^\mathrm{I}}\langle (a^\mathrm{II})^{m_{\mathrm{II},3}}\rangle_{f^\mathrm{II}}
 \delta_{0,\sum_{q_3 \in \mathbf{Q_3}}q_3} \nonumber \\
&+... \nonumber
\end{align}
This  result can be used to devise a similar graphical representation of the averaged Green's function as in Eq. (\ref{G_av_diagram}). After resummation, the self-energy is obtained as the sum over all irreducible diagrams of the form

\begin{align}
\Sigma(k, i\omega)=&
\vcenter{\hbox{
\begin{tikzpicture}
  \begin{feynman}
     \vertex (a) at (0, 0);
     \vertex (i1) at (0, 4)[crossed dot]{};
     \vertex [above=1em of i1] {\(\langle a^\mathrm{I} \rangle_{f^\mathrm{I}}\)};
     \vertex [below=0.5em of a] {\(V^\mathrm{I}_{k,k}\)};
    \diagram*{
    (a) -- [charged scalar, edge label'=\(0\)] (i1),
    };
  \end{feynman}
\end{tikzpicture}}}
+
\vcenter{\hbox{
\begin{tikzpicture}
  \begin{feynman}
     \vertex (a) at (0, 0);
     \vertex (i1) at (0, 4)[crossed dot]{};
     \vertex [above=1em of i1] {\(\langle a^\mathrm{II} \rangle_{f^\mathrm{II}}\)};
     \vertex [below=0.5em of a] {\(V^\mathrm{II}_{k,k}\)};
    \diagram*{
    (a) -- [charged scalar, edge label'=\(0\)] (i1),
    };
  \end{feynman}
\end{tikzpicture}}}
+
\vcenter{\hbox{
\begin{tikzpicture}
  \begin{feynman}
     \vertex (a) at (0, 0);
     \vertex (i1) at (1.5, 4)[crossed dot]{};
     \vertex (b) at (3, 0);
     \vertex [above=1em of i1] {\(\langle a^\mathrm{I} \rangle_{f^\mathrm{I}} \langle a^\mathrm{II} \rangle_{f^\mathrm{II}}\)};
     \vertex [below=0.5em of a] {\(V^\mathrm{I}_{k,k}\)};
     \vertex [below=0.5em of b] {\(V^\mathrm{II}_{k,k}\)};
    \diagram*{
      (b) -- [fermion, edge label=\(q\)] (a) ,
      (a) -- [charged scalar, edge label=\(q-k\)] (i1),
      (b) -- [charged scalar, edge label'=\(k-q\)] (i1),
    };
  \end{feynman}
\end{tikzpicture}}}
+
\vcenter{\hbox{
\begin{tikzpicture}
  \begin{feynman}
     \vertex (a) at (0, 0);
     \vertex (i1) at (1.5, 4)[crossed dot]{};
     \vertex (b) at (3, 0);
     \vertex [above=1em of i1] {\(\langle a^\mathrm{II} \rangle_{f^\mathrm{II}} \langle a^\mathrm{I} \rangle_{f^\mathrm{I}}\)};
     \vertex [below=0.5em of a] {\(V^\mathrm{II}_{k,k}\)};
     \vertex [below=0.5em of b] {\(V^\mathrm{I}_{k,k}\)};
    \diagram*{
      (b) -- [fermion, edge label=\(q\)] (a) ,
      (a) -- [charged scalar, edge label=\(q-k\)] (i1),
      (b) -- [charged scalar, edge label'=\(k-q\)] (i1),
    };
  \end{feynman}
\end{tikzpicture}}}
\nonumber \\
&
+
\vcenter{\hbox{
\begin{tikzpicture}
  \begin{feynman}
     \vertex (a) at (0, 0);
     \vertex (i1) at (1.5, 4)[crossed dot]{};
     \vertex (b) at (3, 0);
     \vertex [above=1em of i1] {\(\langle (a^\mathrm{I})^2 \rangle_{f^\mathrm{I}} \)};
     \vertex [below=0.5em of a] {\(V^\mathrm{I}_{k,k}\)};
     \vertex [below=0.5em of b] {\(V^\mathrm{I}_{k,k}\)};
    \diagram*{
      (b) -- [fermion, edge label=\(q\)] (a) ,
      (a) -- [charged scalar, edge label=\(q-k\)] (i1),
      (b) -- [charged scalar, edge label'=\(k-q\)] (i1),
    };
  \end{feynman}
\end{tikzpicture}}}
+
\vcenter{\hbox{
\begin{tikzpicture}
  \begin{feynman}
     \vertex (a) at (0, 0);
     \vertex (i1) at (1.5, 4)[crossed dot]{};
     \vertex (b) at (3, 0);
     \vertex [above=1em of i1] {\(\langle (a^\mathrm{II})^2 \rangle_{f^\mathrm{II}} \)};
     \vertex [below=0.5em of a] {\(V^\mathrm{II}_{k,k}\)};
     \vertex [below=0.5em of b] {\(V^\mathrm{II}_{k,k}\)};
    \diagram*{
      (b) -- [fermion, edge label=\(q\)] (a) ,
      (a) -- [charged scalar, edge label=\(q-k\)] (i1),
      (b) -- [charged scalar, edge label'=\(k-q\)] (i1),
    };
  \end{feynman}
\end{tikzpicture}}}
+
\vcenter{\hbox{
\begin{tikzpicture}
  \begin{feynman}
     \vertex (a) at (0, 0);
     \vertex (i1) at (1.5, 4)[crossed dot]{};
     \vertex (i2) at (1.5, 2)[crossed dot]{};
     \vertex (b) at (1.5, 0);
     \vertex (c) at (3, 0);
     \vertex [above = 1em of i1] {\(\langle (a^\mathrm{I})^2 \rangle_{f^\mathrm{I}}\)};
     \vertex [above=1em of i2] {\(\langle a^\mathrm{II} \rangle_{f^\mathrm{II}}\)};
     \vertex [below=0.5em of a] {\(V^\mathrm{I}_{k,k}\)};
     \vertex [below=0.5em of b] {\(V^\mathrm{II}_{k,k}\)};
     \vertex [below=0.5em of c] {\(V^\mathrm{I}_{k,k}\)};
    \diagram*{
      (b) -- [fermion, edge label=\(q\)] (a) ,
      (a) -- [charged scalar, edge label=\(q-k\)] (i1),
      (b) -- [charged scalar, edge label'=\(0\)] (i2),
      (c) -- [charged scalar, edge label'=\(k-q\)] (i1),
      (c) -- [fermion, edge label=\(q\)] (b) ,
    };
  \end{feynman}
\end{tikzpicture}}}
+..., \label{se_mixed} 
\end{align}
which obey simple Feynman rules again. The solid-lined propagators with momentum $k$ denote a matrix-valued free GF $G^0(k, i \omega)$. The dashed propagators carry an also matrix-valued factor $V^\mathrm{I}_{q_\mathrm{L},q_\mathrm{R}}$ or $V^\mathrm{II}_{q_\mathrm{L},q_\mathrm{R}}$, where $q_\mathrm{L}$ is the momentum leaving the vertex of the dashed and the two solid propagators to the left and $q_\mathrm{R}$ the momentum joining it from the right. A vertex of $m_{\mathrm{I}}$ dashed propagators with  $V^\mathrm{I}_{q_\mathrm{L},q_\mathrm{R}}$ 
and $m_{\mathrm{II}}$ dashed propagators with  $V^\mathrm{II}_{q_\mathrm{L},q_\mathrm{R}}$ obtains  a prefactor of 
$\langle(a^\mathrm{I})^{m_{\mathrm{I}}}\rangle_{f^\mathrm{I}}  \langle(a^\mathrm{II})^{m_{\mathrm{II}}}\rangle_{f^\mathrm{II}}$ . The dashed propagators formally carry the momentum $q_\mathrm{R} - q_\mathrm{L}$ and all momenta joining a vertex of multiple dashed propagators add up to zero.  A sum $\frac{1}{N_\mathrm{sites}}\sum_q$ over all momenta inside a closed loop is implied. \\

It is only necessary to consider distributions $f$ with vanishing first moment, as we can always redefine 

\begin{align}
H =& H_0 + V = \sum_k c^\dagger_k H_0(k) c_k +  \sum_{j= 1} ^ {N_\mathrm{site}} \sum_{k,k'} \frac{e^{-i j (k- k')}}{N_\mathrm{sites}} c_k^\dagger \left(a^\mathrm{I}_j V^\mathrm{I}_{k,k'} + a^\mathrm{II}_j V^\mathrm{II}_{k,k'} \right) c_{k'}. \nonumber \\
=& \sum_k c^\dagger_k \underbrace{\left ( H_0(k) + \langle a^\mathrm{I} \rangle_{f^\mathrm{I}} V^\mathrm{I}_{k,k} 
+ \langle a^\mathrm{II} \rangle_{f^\mathrm{II}} V^\mathrm{II}_{k,k} \right )}_{H^{\mathrm{new}}_0} c_k 
+  \sum_{j= 1} ^ {N_\mathrm{site}} \sum_{k,k'} \frac{e^{-i j (k- k')}}{N_\mathrm{sites}} c_k^\dagger \left (a^\mathrm{I}_j V^\mathrm{I}_{k,k'} + a^\mathrm{II}_j V^\mathrm{II}_{k,k'} \right ) c_{k'} \nonumber \\
&- \sum_{k,k'} \delta_{k,k'} c^\dagger_k \left ( \langle a^\mathrm{I} \rangle_{f^\mathrm{I}} V^\mathrm{I}_{k,k'} + \langle a^\mathrm{II} \rangle_{f^\mathrm{II}} V^\mathrm{II}_{k,k'} \right ) c_{k'} \nonumber \\
=& \sum_k c^\dagger_k H^{\mathrm{new}}_0c_k 
+  \sum_{j= 1} ^ {N_\mathrm{site}} \sum_{k,k'} \frac{e^{-i j (k- k')}}{N_\mathrm{sites}} c_k^\dagger \left( \left (a^\mathrm{I}_j 
- \langle a^\mathrm{I} \rangle_{f^\mathrm{I}} \right) V^\mathrm{I}_{k,k'} 
+ \left( a^\mathrm{II}_j -  \langle a^\mathrm{II} \rangle_{f^\mathrm{II}}\right) V^\mathrm{II}_{k,k'}\right) c_{k'}. \nonumber 
\end{align}
Then, the self-energy from Eq. (\ref{se_mixed}) is apparently given by the sum of the self energies emerging from the perturbations  $V^\mathrm{I}_{k,k'}$ and $V^\mathrm{II}_{k,k'}$ alone plus mixed terms
starting at fourth order ($\propto \langle (a^\mathrm{I})^2 \rangle_{f^\mathrm{I}} \langle (a^\mathrm{II})^2 \rangle_{f^\mathrm{II}}$).\\

As was shown in the first part of this section, we require $OS$ and $NN$ terms in a single perturbation to create an $i \sigma_z$ contribution in the FBA self-energy correction to our chirally symmetric two-banded model.
In conclusion, if we split this perturbation into two uncorrelated perturbations with only $OS$ and only $NN$ terms, an $i\sigma_z$-term cannot emerge in FBA. We could only hope for EPs from higher-order effects, which is unlikely and beyond the scope of this analysis anyway. This also agrees with the results from the full numerics, where EPs could only be observed in systems with correlated $OS$ and $NN$ terms in the perturbation.

\end{document}